\documentclass[11pt]{article}
\usepackage[sort&compress,square,comma,numbers]{natbib}

\usepackage[labelfont=bf]{caption} % Bold figure number
\usepackage[font={small}]{caption}   %% FIGURE CAPTION FONT MODIFICATION

\usepackage{subcaption}
\usepackage{amsmath}
\usepackage{amssymb}
\usepackage{url}
\usepackage{placeins}
\usepackage{textcomp}
\usepackage{graphicx}
\usepackage{float} % Force Figure placement (H)
\usepackage{chngpage}

\usepackage[usenames,dvipsnames]{color}    % ADDED FOR COLOR NAMES

\usepackage[usenames,dvipsnames]{color}    % ADDED FOR COLOR NAMES
\usepackage{soul}   % SOUL Added for underlines
\setstcolor{red} % strikethrough color
\setul{}{1.5pt}  % line thickness

\title{\textbf{Breaking time-reversal symmetry with acoustic pumping of nanophotonic circuits}}
\author{Donggyu B Sohn, Seunghwi Kim, Gaurav Bahl$^{\ast}$\\
	\\
	\footnotesize{Mechanical Science and Engineering, University of Illinois at Urbana-Champaign,}\\
	\footnotesize{Urbana, Illinois, USA}\\
	\footnotesize{$^\ast$To whom correspondence should be addressed; E-mail: bahl@illinois.edu}
	}

\date{}

\begin{document}
\maketitle

\begin{abstract}
Achieving non-reciprocal light propagation via stimuli that break time-reversal symmetry, without magneto-optics, remains a major challenge for integrated nanophotonic devices. Recently, optomechanical microsystems in which light and vibrational modes are coupled through ponderomotive forces, have demonstrated strong non-reciprocal effects through a variety of techniques, but always using optical pumping. None of these approaches have demonstrated bandwidth exceeding that of the mechanical system, and all of them require optical power, which are both fundamental and practical issues. Here we resolve both of these challenges through breaking of time-reversal symmetry using an acoustic pump in an integrated nanophotonic circuit. GHz-bandwidth optomechanical non-reciprocity is demonstrated using the action of a 2-dimensional surface acoustic wave pump, that simultaneously provides non-zero overlap integral for light-sound interaction and also satisfies the necessary phase-matching. We use this technique to produce a simple frequency shifting isolator (i.e. a non-reciprocal modulator) by means of indirect interband scattering. We demonstrate mode conversion asymmetry up to 15 dB, efficiency as high as 17\%, over bandwidth exceeding 1 GHz.
\end{abstract}

% ==================================
%% Temporarily commented this out to work on the rest
% ==================================
%\section{Introduction}

	Non-reciprocal devices, in which time reversal symmetry is broken for light propagation, provide critical functionalities for signal routing and source protection in photonic systems. The most commonly encountered non-reciprocal devices are isolators and circulators, which can be implemented using a variety of techniques encompassing magneto-optics~\cite{Huang:17,Ross:11}, parity-time symmetry breaking~\cite{Peng2014}, spin-polarized atom-photon interactions~\cite{Sayrin:15,Scheucher:16}, and optomechanical interactions~\cite{KangM.2011,Kim2016,Poulton2012,Shen2016,Fang:2017,Ruesink:2016,Dong2015}. On the other hand, recent developments reveal a much broader and compelling vision of using time-reversal symmetry breaking for imparting protection against defects, through analogues of the quantum Hall effect~\cite{Halperin:82} in both topological~\cite{Wang:09,Hafezi:13,Susstrunk:15} and non-topological systems~\cite{Kim2017}. 

	The use of optomechanical coupling~\cite{RevModPhys.86.1391} for breaking time-reversal symmetry via momentum biasing~\cite{Kim2015,Dong2015} and synthetic magnetism~\cite{Fang:2017,Ruesink:2016} is particularly attractive since strong dispersive features can be readily produced, without needing materials with gain or magneto-optical activity. Additionally, the potential for complete isolation with ultralow loss~\cite{Kim2016} is a significant advantage over state-of-the-art in chip scale magneto-optics. All these systems feature dynamic reconfigurability through the pump laser fields and can potentially be implemented in foundries with minimal process modification. Unfortunately, all realizations of optomechanical non-reciprocal interactions to date only operate over kHz-MHz bandwidth. This fundamental constraint arises simply because the interaction is determined by the mechanical linewidth, which is traditionally 6-9 orders of magnitude lower than the optical system (potentially several THz). In this work we present a new approach for optomechanical non-reciprocity where the bandwidth of the effect is no longer determined by the mechanics, but is instead determined by the photonic modes. We achieve this by inverting the roles of the mechanical and optical modes in a common optomechanical configuration so that acoustic pumping, as opposed to optical pumping, is used to break time-reversal symmetry. The practical implications of this new pumping strategy are transformative; we no longer need any additional lasers to drive the system, and more importantly, the linearity of the non-reciprocal effect is no longer limited to small optical signals. For the first time, an integrated nanophotonic device is produced that exhibits broken time-reversal symmetry over GHz bandwidth using a phonon pump.

\vspace{12pt}

	The specific device that we implement for this demonstration is a non-reciprocal nanophotonic modulator. Light entering the device from one direction is transferred to a different optical band through phonon-mediated momentum shift and energy shift, i.e. indirect interband scattering~\cite{Kuhn:71,Hwang:97,Yu2009}. Light entering from the opposite direction is simply resonantly dissipated. As such, the presented device operates as a frequency shifting optical isolator exhibiting 15 dB of contrast and up to 17\% mode conversion efficiency. Unlike sophisticated electro-optic implementation of this idea~\cite{lira2012}, acoustic phonons naturally provide large momentum shifts at practical driving frequencies. Thus, it is unnecessary to slow down the effective phase velocity of the pumping signal to achieve indirect interband scattering using a phononic pump. The acoustic method also circumvents free carrier absorption that otherwise generates large losses in electro-optic waveguides.

\begin{figure}[!hp]
	\begin{adjustwidth}{-1in}{-1in}
		\makebox[\textwidth][c]{\includegraphics[width=1.3\textwidth]{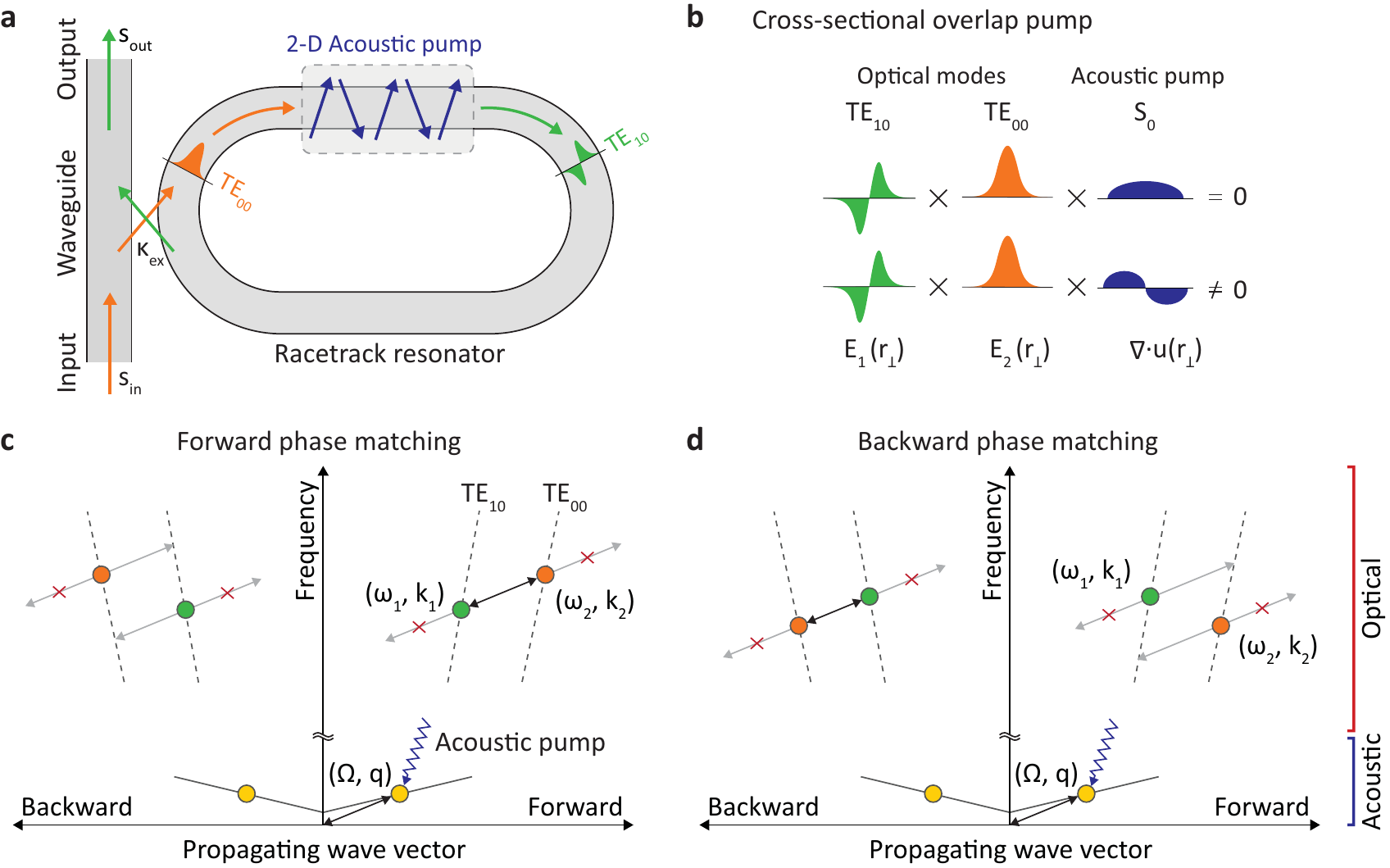}}
		\centering
		\caption{
			(a) Conceptual schematic of acoustically pumped non-reciprocal nanophotonic modulator. The device is composed of a racetrack resonator supporting two optical modes (TE$_{10}$ and TE$_{00}$). An electrically driven 2-dimensional acoustic wave (pump) simultaneously breaks orthogonality between the optical modes while also satisfying the phase matching condition. $s_{in}$ and $s_{out}$ represent input and output signals from the waveguide. 
			(b) Illustration of the transverse mode profiles ($E_1 \left( \textbf{r}_\perp \right)$, $E_2 \left( \textbf{r}_\perp \right)$ are electric fields, $u\left( \textbf{r}_\perp \right)$ is material displacement) shows the odd acoustic pump profile needed to obtain non-zero overlap integral. 
			(c) The required phase matching condition illustrated in frequency-momentum space. The acoustic pump is launched in the forward direction with frequency and momentum ($\Omega,q$). The lower momentum optical mode (TE$_{10}$) has frequency and momentum ($\omega_1,k_1$) and the higher momentum optical mode (TE$_{00}$) has frequency and momentum ($\omega_2,k_2$). When the resonance frequency of the TE$_{00}$ mode is higher than of the TE$_{10}$ mode, the phase matching condition can be satisfied in the forward direction. 
			(d) Conversely, when the resonance frequency of the TE$_{10}$ mode is higher than the TE$_{00}$ mode, the phase matching condition is satisfied in the backward direction.
		}
		\label{fig1:PM}
	\end{adjustwidth}
\end{figure}

To explain this approach qualitatively, we consider an optomechanically-active racetrack resonator ~\cite{Shin2013,Kitt2016} supporting quasi-TE$_{10}$ ($\omega_1, k_1$) and quasi-TE$_{00}$ ($\omega_2, k_2$) as visualized in Fig.~\ref{fig1:PM}. For convenience we drop the 'quasi-' prefix. Indirect intermodal scattering~\cite{Yu2009} can be enabled between the optical modes as a result of the photoelastic perturbations of the medium by the driven acoustic wave ($\Omega, q$) \cite{Hwang:97,Kuhn:71}.
While this optically resonant structure sacrifices the bandwidth over which acousto-optical interactions can occur, it provides giant opto-acoustic gain that is necessary to obtain appreciable light-vibration coupling within a small form factor \cite{Dostart:16}.
The requisite phase matching conditions are illustrated in $\omega-k$ space in Fig.~\ref{fig1:PM}c,d. Under normal conditions, the TE$_{00}$ momentum is higher than that of the TE$_{10}$ mode (i.e. $k_2 > k_1$). 

We first consider the case illustrated in Fig.~\ref{fig1:PM}c where the resonance frequency of the TE$_{00}$ racetrack mode is higher than that of the TE$_{10}$ mode ($\omega_2 > \omega_1$).  Here, a carefully designed acoustic transducer that generates phonons ($\Omega, q$) between the modes can help satisfy the phase matching condition ($\Omega=\omega_2-\omega_1$, $q = k_2-k_1$) for acousto-optical scattering for forward propagating optical signals. For light propagating in the opposite direction, the momentum difference between the optical modes is now ($\Omega, -q$) which is not satisfied by the driven phonons. Thus the system exhibits broken time-reversal symmetry, i.e. intermodal scattering is permitted only for forward signals while the backward signals see no such effect. We can now also consider the case shown in Fig.~\ref{fig1:PM}d where the resonance frequency relation is opposite. In this case the phase-matching of scattering between the optical modes requires phonons having ($\Omega, -q$) in the forward direction and ($\Omega,q$) in the backward direction. Thus, forward propagating phonons can only phase-match backward propagating optical modes in this case. 

In addition to the above phase matching requirement, the acoustic wave must also assist with breaking orthogonality.  
When both optical modes are of TE polarization, the intermodal optomechanical coupling coefficient ($\beta$) is proportional to the cross-sectional overlap integral of the optical modes $E_1 \left( \textbf{r}_\perp \right)$, $ E_2 \left( \textbf{r}_\perp \right)$ and the acoustic mode displacement $u\left( \textbf{r}_\perp \right)$ given by~\cite{Agrawal2013}: 	
\begin{align}
\beta \propto \iint E_1 \left( \textbf{r}_\perp \right) E_2 \left( \textbf{r}_\perp \right) \left(\nabla\cdot u\left( \textbf{r}_\perp \right) \right) d^2\textbf{r}_\perp
\end{align}
As shown in Fig.~\ref{fig1:PM}b, the electric fields of TE$_{10}$ and TE$_{00}$ modes have odd and even shapes in the transverse direction, respectively. Therefore, the density variation associated with the acoustic wave must be asymmetric with respect to the center of the waveguide to ensure non-zero $\beta$. In the case where the node is located exactly at the center of the waveguide, we can maximize the intermodal coupling and simultaneously suppress intramodal scattering by balancing out compressive and tensile strain in the waveguide.

The acousto-optic interaction in the forward phase matched case (Fig.~\ref{fig1:PM}c) can now be described using the coupled equations of motion for the optical fields (backward phase matched case in Supplement \S S3):
\begin{align}
\dfrac{\partial}{\partial t} 
\begin{pmatrix}
a_1\\
a_2 
\end{pmatrix}
=
-i
\begin{pmatrix}
\omega_1-i\kappa_1/2 & G_{ph}^*e^{i\Omega t}\\
G_{ph}e^{-i\Omega t}& \omega_2-i\kappa_2/2
\end{pmatrix}
\begin{pmatrix}
a_1\\
a_2 
\end{pmatrix}
+
\begin{pmatrix}
\sqrt{\kappa_{ex1}}\\
\sqrt{\kappa_{ex2}}
\end{pmatrix}
s_{in}e^{-i\omega_l t}
\end{align}
\noindent where $a_1$ ($a_2$) is the intracavity field, $\kappa_1$ ($\kappa_2$) is the loaded decay rate of the TE$_{10}$ (TE$_{00}$) mode, $G_{ph}=\beta u$ is the phonon-enhanced optomechanical coupling rate, and $u$ is displacement associated with the acoustic pump. 
%Here, $\beta$ is inter-modal optomechanical coupling coefficient and 
%
Here we assume that an input field $s_{in}$ at carrier frequency $\omega_l$ is provided to the resonator via an evanescently coupled waveguide (Fig.~\ref{fig1:PM}a) with coupling rates $\kappa_{ex1}$ and $\kappa_{ex2}$ to the TE$_{10}$ and TE$_{00}$ optical racetrack modes respectively.
In the case where we probe the $a_2$ mode, we can express the optical susceptibility as $\chi_{om}(\omega)=\left[\kappa_2/2+i(\omega_2-\omega)+\alpha(\omega)\right]^{-1}$
where $\omega$ is Fourier frequency and $\alpha(\omega)=\left| G_{ph} \right|^2/ \left[ \kappa_1/2 + i(\omega_1+\Omega-\omega) \right]$ is an additional optical loss rate induced by the acousto-optic interaction. 
In contrast, optically pumped optomechanical systems~\cite{KangM.2011,Poulton2012,Shen2016,Fang:2017,Ruesink:2016,Dong2015,Kim2017,RevModPhys.86.1391,Kim2015,Kim2016} have an interaction term of the form  $\alpha(\omega)=\left| G \right|^2/ \left[ \Gamma/2 + i(\Omega_m+\omega_p-\omega) \right]$ where $G=\beta a_1$ is the photon-enhanced optomechanical coupling rate, $\Gamma$ is decay rate of the mechanical mode, $\Omega_m$ is the mechanical resonant frequency, and $\omega_p$ is the pump laser frequency~\cite{Kim2016}. We can immediately see that the interaction bandwidth in the acoustically pumped case is no longer defined by the linewidth $\Gamma$ of the mechanical mode, but instead by the linewidth $\kappa_1$ of the $a_1$ optical mode. This feature enables an orders-of-magnitude higher bandwidth non-reciprocal interaction.

	 \vspace{12pt}

%=========================
%\section{Device design}
%=========================
%\vspace{24pt}

\begin{figure}[t!]
	\begin{adjustwidth}{-1in}{-1in}
 		\makebox[\textwidth][c]{\includegraphics[width=1.2\textwidth]{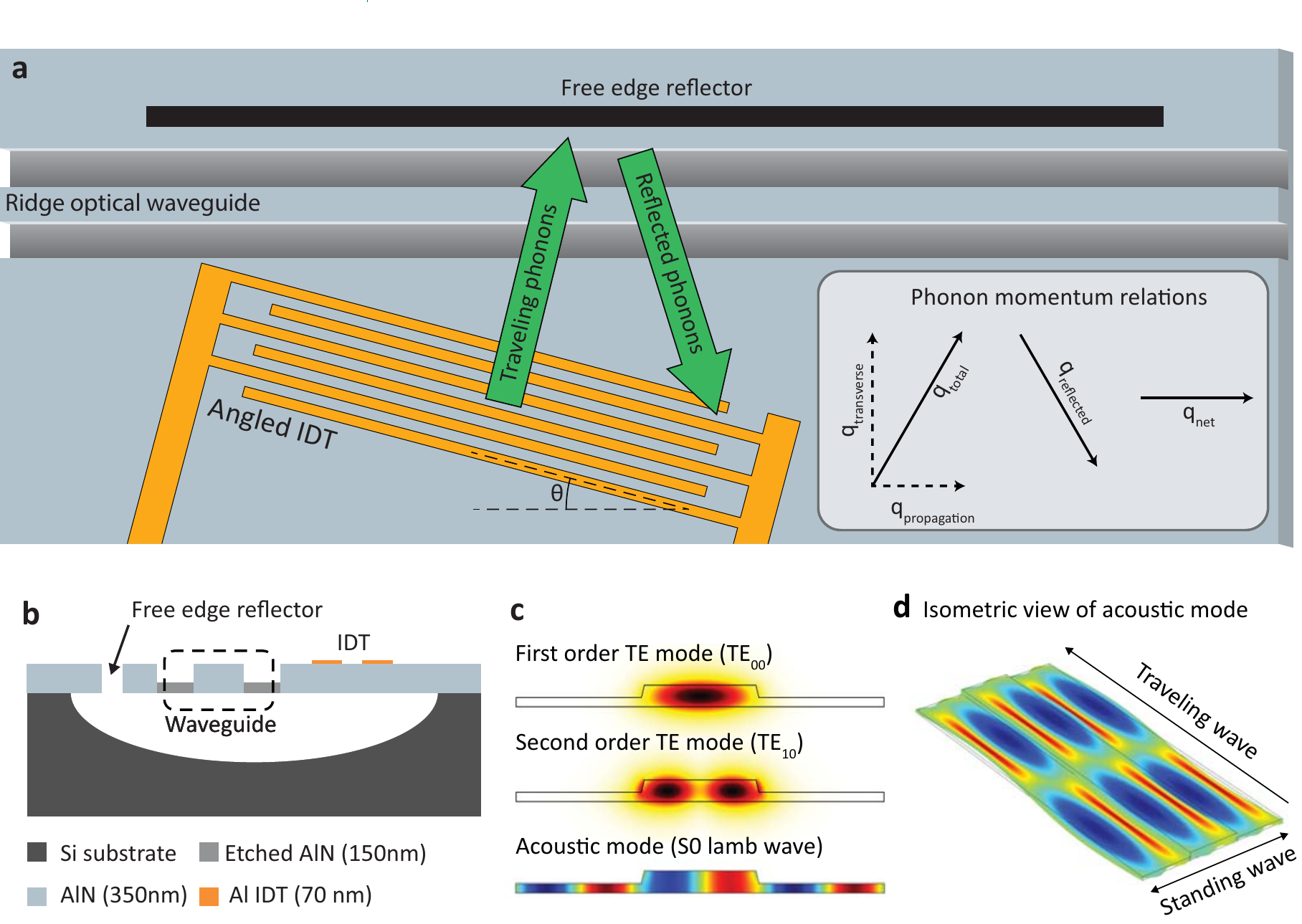}}
 		\centering
 		\caption{
 			(a) Schematic of the phonon-photon interaction region. A 2-dimensional acoustic wave is generated using an angled interdigitated transducer (IDT) that provides momentum in both transverse and propagating directions. The pitch of the IDT ($\Lambda$ = 546 nm) determines total momentum of driven phonons, while the angle of the IDT ($\theta$ = 7.06$^{\circ}$) determines the ratio between transverse and propagating phonon momenta. The free edge reflector is situated such that a standing acoustic wave is formed in the transverse direction and its node is placed in the middle of the nanophotonic waveguide. (b) Cross-section schematic of the phonon photon interaction region. (c) FEM simulated mode shapes of the TE$_{10}$ and TE$_{00}$ optical modes and the S$_0$ acoustic wave. (d) Isometric view of the 2-dimensional acoustic wave propagating along the optical waveguide.
 		}
 		\label{fig2:a}
	\end{adjustwidth}
 \end{figure}

\begin{figure}[t!]
	\begin{adjustwidth}{-1in}{-1in}
		\makebox[\textwidth][c]{\includegraphics[width=1.2\textwidth]{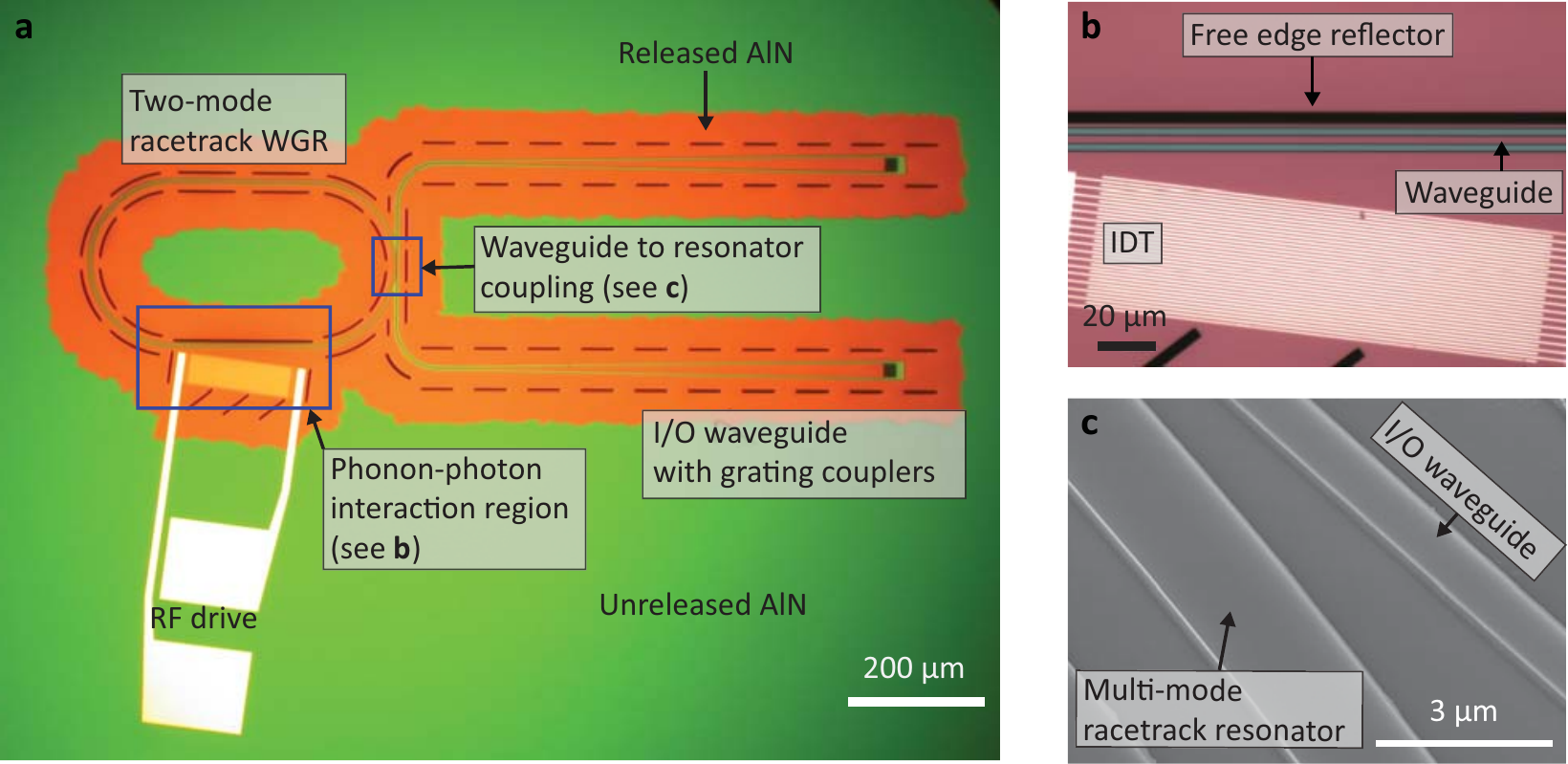}}
		 		\centering
		\caption{(a) True color micrograph of the acoustically pumped non-reciprocal nanophotonic modulator. (b) Close up of the phonon-photon interaction region showing the interdigitated transducer (IDT) and the free edge acoustic reflector. (c) SEM image of the 2.2 $\mu$m width racetrack resonator and the 800 nm single mode waveguide in the evanescent coupling region.}
		\label{fig3:img}
	\end{adjustwidth}
\end{figure}

\begin{figure}[!ht]
	\begin{adjustwidth}{-1in}{-1in}
		\makebox[\textwidth][c]{\includegraphics[width=1.3\textwidth]{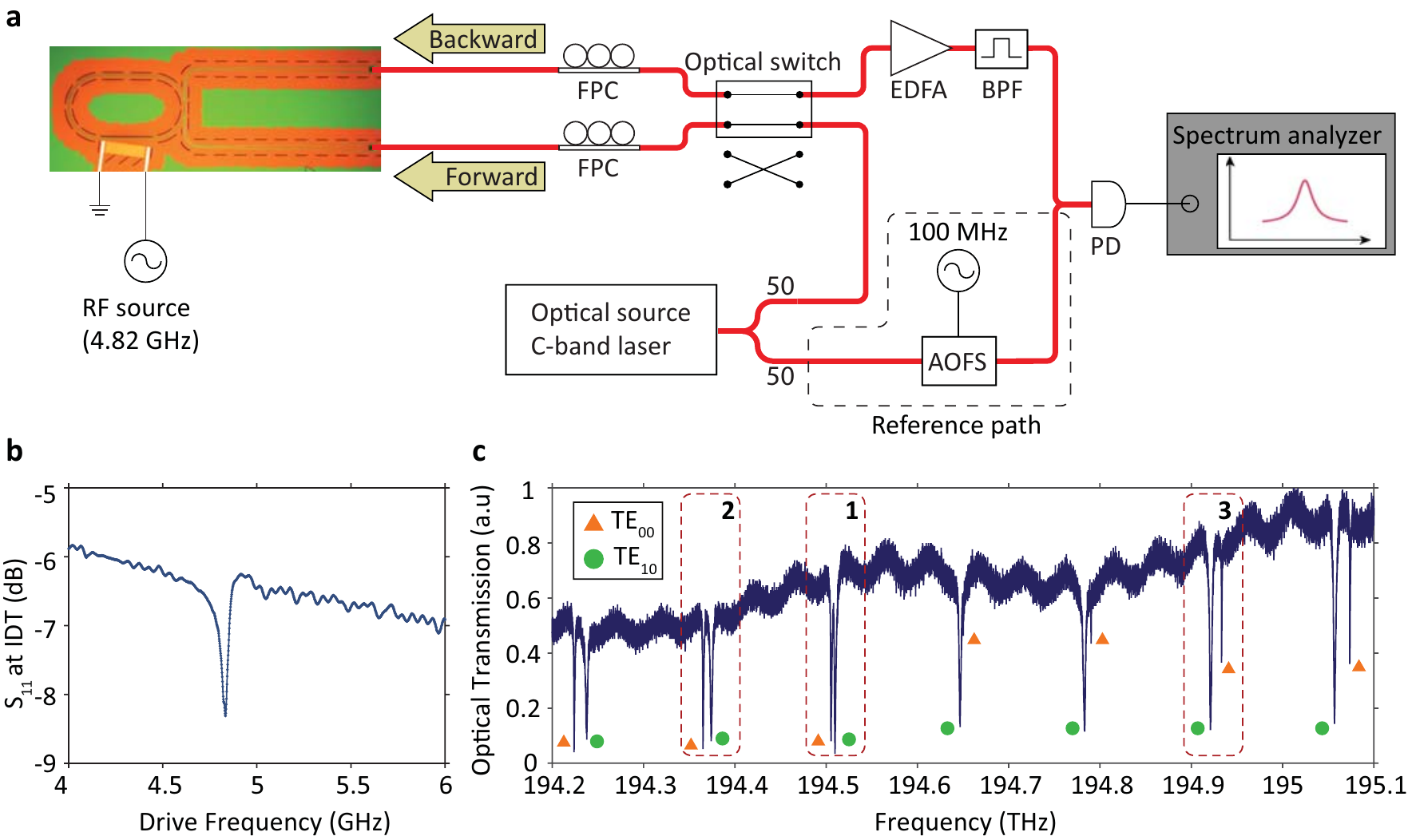}}
		 		\centering
		\caption{(a) Measurement setup. Light from a tunable external cavity diode laser is split with a 50:50 coupler to prepare a device probing path and a reference path. An acousto-optic frequency shifter (AOFS) offsets the reference path by 100 MHz to enable heterodyne detection via a high speed photodetector (PD). An off-chip optical switch controls the directionality of the light entering the on-chip waveguide. The light coming out from the device is amplified using an erbium-doped fiber amplifier (EDFA) to compensate the loss from the grating coupler. A tunable band pass filter (BPF) is placed to filter out the extra noise produced by the EDFA. (b) Measured reflection coefficient (s$_{11}$ response) of the IDT using a vector network analyzer. (c) Measured optical transmission without acoustic drive. Triangles and circles represent optical modes of TE$_{00}$ and TE$_{10}$ families respectively. Mode pairs 1,2, and 3 are used for the intermodal scattering experiment (Fig.~5). 
		} 
		\label{fig4:mea}
	\end{adjustwidth}
\end{figure}

We fabricate the nonreciprocal modulator on an aluminum nitride (AlN) device layer on air platform, having an underlying a silicon handle wafer (Fig.~\ref{fig2:a}a,b). This selection of materials ensures CMOS compatibility~\cite{Xiong:12}. Here, the AlN supports the optical modes due to its high transparency in the telecom band~\cite{Xiong:121,Xiong:12} and additionally functions as an excellent acoustic material on which phonons can be piezoelectrically driven (Fig.~\ref{fig2:a}c,d) ~\cite{Tadesse2014,Li:15,piazza:06}. Details on the fabrication process are provided in Methods below.
The device is composed of an AlN racetrack resonator that supports the required TE$_{00}$ and TE$_{10}$ modes around 1550 nm using a wrapped ridge waveguide (details provided in Supplement~\S S7). Other optical modes are suppressed by limiting the width (2.2 $\mu$m) and thickness (350 nm) of the racetrack waveguide. 
%
%\comment{GB: Now we explain the adjacent linear waveguide that is used for coupling to the resonator, and provide its geometry \Ra} 
The optical modes of the racetrack are accessed through an adjacent linear single-mode waveguide with a width of 800~nm that is coupled evanescently to the resonator at a single point. Grating couplers at the ends of the linear waveguide are used to provide optical access to the system.
The as-fabricated waveguide geometries were evaluated using electron microscopy, which permitted more accurate refinement and finite-element simulation of the optical modes in Comsol Multiphysics. This procedure also allows us to evaluate the material refractive index as $(n_{AlN}=2.07)$ by matching against experimental measurement of the free spectral range (FSR) of each optical mode family within the resonator.
The frequency difference between pairs of optical resonances varies due to the distinct dispersion of the TE$_{00}$ and TE$_{10}$ mode families (Fig.~\ref{fig4:mea}c).
The acoustic pump is provided to the resonator using an interdigitated transducer (IDT) that is fabricated on the same piezoelectric AlN substrate. %\st{ as the resonator.}
The IDT pitch and angle are selected carefully in order to generate a 2-dimensional acoustic wave having the correct momentum in both propagating and transverse directions, as shown in Fig.~\ref{fig2:a}a. 
In order to satisfy momentum conservation in the propagating direction, an acoustic propagation constant of $q_{propagating} = 3.54 \times 10^5 m^{-1}$ is required.
Since we simultaneously require a standing wave in the transverse direction, an acoustic free edge reflector is fabricated by cutting the AlN device layer through to the air below (Fig.~\ref{fig2:a}a,b). This free edge reflector is placed at $2\lambda$ away from the waveguide in order to situate an acoustic node at the center of the waveguide and obtain an odd transverse profile (Fig.~\ref{fig1:PM}b).
Based on simulation, the cross-sectional overlap integral $\beta$ is maximized when the transverse acoustic wavelength and the width of the optical wave guide are matched, setting transverse propagation constant to $q_{transverse} = 2\pi / 2.2~\mu m^{-1} = 2.86 \times 10^6 m^{-1}$. 
Accounting for the transverse and propagating components of the acoustic wave, the total wave vector of the acoustic wave launched by the IDT is calculated as $q_{total} = \sqrt{ q_{propagating}^2 + q_{transverse}^2 } \approx 2.88 \times 10^6 m^{-1}$. 
The IDT angle is then set to $\theta=\tan^{-1}(q_{propagating}/q_{transverse}) = 7.06 ^{\circ} $ , and pitch to $\Lambda=\pi / 2 q_{total}$ = 546 nm. 
The required driving frequency of 4.82 GHz is calculated using finite element simulation of the S$_0$ Lamb surface acoustic wave dispersion for the selected propagation constant.
Fig.~\ref{fig2:a}d presents a simulation of the 2-dimensional acoustic mode shape in the racetrack waveguide region. 
The intermodal modulation frequency and momentum can be tailored for different phononic and photonic modes by simply modifying the IDT parameters, without changing material phonon dispersion.
Micrographs of the fabricated device are presented in Fig.~\ref{fig3:img}.

\vspace{12pt}

We experimentally demonstrate non-reciprocal modulation within the system by measuring the optical sidebands for both forward and backward probe signals, using the measurement setup shown in Fig.~\ref{fig4:mea}a (see also Methods). 
The primary acoustic component of the system, i.e. the IDT, is first characterized using s-parameter measurement (Fig.~\ref{fig4:mea}b) using a calibrated RF probe by means of an electronic vector network analyzer. 
The measured reflection coefficient (s$_{11}$ parameter) shows a resonant dip corresponding to efficient conversion of the input electronic stimulus into the acoustic wave. We can directly measure this characteristic acoustic resonance at 4.82 GHz, corresponding to the S$_0$ Lamb surface acoustic wave on the AlN substrate. 
In Fig.\ref{fig4:mea}c we present the measured optical transmission spectrum from the perspective of the coupling waveguide, where the transmission dips corresponding to the modes of the racetrack resonator are clearly visible. The TE$_{00}$ mode is seen to have higher Q factor (Q$_{\textrm{TE}_{00}\textrm{,loaded}} \approx$ 170,000) than the TE$_{10}$ mode (Q$_{\textrm{TE}_{10}\textrm{,loaded}} \approx$ 102,000). 
The measured free spectral ranges (FSR) of TE$_{00}$ and TE$_{10}$ resonances are respectively 140.5 GHz and 136 GHz near 194.7 THz optical frequency. Therefore, the inter-modal frequency difference changes by approximately 4.5 GHz for each consecutive mode pair. On the other hand, the momentum difference of each mode pair is the same since the azimuthal mode order of the pairs is the same.

\vspace{12pt}

\begin{figure}[!ht]
	\begin{adjustwidth}{-1in}{-1in}
		\makebox[\textwidth][c]{\includegraphics[width=1.15\textwidth]{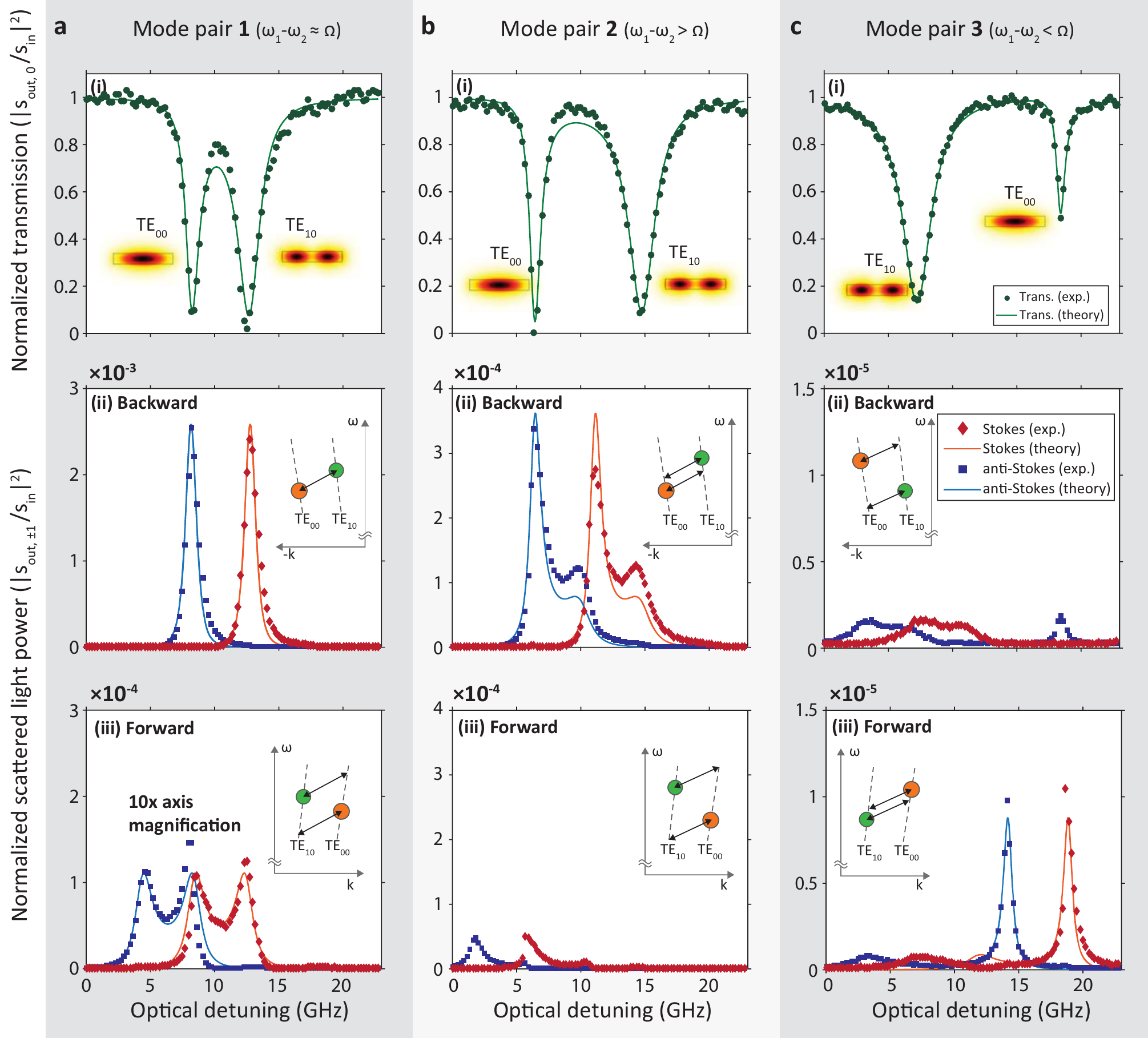}}
		\centering
		\caption{
			Experimental demonstration of non-reciprocal optomechanical modulation. Three cases corresponding to the optical mode pairs indicated in Fig.\ref{fig4:mea}c are presented. 
			(a) Mode pair 1 corresponds to a near-perfect backward phase-matching condition. Here the optical modes are separated by $\omega_2 - \omega_1 = 4.55$ GHz (TE$_{00}$ mode is located at lower frequency) such that a 4.82 GHz acoustic mode has the correct momentum to enable phase matching.
			(b) Mode pair 2 corresponds to $\omega_2-\omega_1 = 8.01$ GHz corresponding to imperfect phase matching. (c) Mode pair 3 ($\omega_2-\omega_1 = -11.76$ GHz) has the TE$_{00}$ mode located at higher frequency; thus phase matching is satisfied in the forward direction. 
			The top row (i) presents the transmitted signal at the optical carrier frequency component (Green), the second row (ii) presents measurements of Stokes (Red) and anti-Stokes (Blue) sidebands when the system is probed in the backward direction, and the third row (iii) presents sideband measurements for forward probing.
			Legends corresponding to all figures are presented in column (c). 
		}
		\label{fig5:is}
	\end{adjustwidth}
\end{figure}

Experimental measurements of the interband mode conversion are presented in Fig.~\ref{fig5:is}. 
We consider three cases corresponding to mode pairs 1, 2, 3 as marked in Fig.~\ref{fig4:mea}c -- which serve to illustrate both perfect and imperfect phase matching situations within the system. As mentioned above, these mode pairs all have identical separation in momentum-space.
In this experiment, the RF frequency is fixed to 4.82 GHz where IDT can most efficiently actuate the acoustic wave.
We sweep the optical probe across each pair of modes while measuring the power of transmitted carrier frequency component ($s_{out,0}$), down-converted Stokes sideband ($s_{out,-1}$), and up-converted anti-Stokes sideband ($s_{out,+1}$) (Fig.~\ref{fig5:is}i, ii, iii) simultaneously. The measured power is normalized against the input power to the waveguide ($s_{in}$).
The RF drive power is set to 0 dBm so that the optical sidebands are small compared to the input light.

We first examine sideband generation and reciprocity in the case where both phase matching and the frequency matching are well satisfied (Fig. 5a). 
Here, the optical resonance frequency of the TE$_{00}$ mode is lower than the resonance frequency of the TE$_{10}$ optical mode by 4.55 GHz, implying that the phase matching condition is satisfied in the backward direction for a forward directed phonon pump (as illustrated in Fig.~\ref{fig1:PM}d). 
Measurements show (Fig.~\ref{fig5:is}a,ii) that when each optical mode is probed in the backward direction, resonant sideband generation occurs with the assistance of the second optical mode.
For laser detuning between 5-10 GHz (arbitrary reference) light from the waveguide primarily enters TE$_{00}$ mode resulting in a strong anti-Stokes sideband. Scattering to Stokes is strongly suppressed (45 dB smaller than the anti-Stokes) since there is no optical mode available in the resonator. 
%
%The difference between the power of the Stokes and anti-Stokes is ~45 dB when the carrier light is fixed on the resonance. 
%
Similarly, for laser detuning between 10-15 GHz, light from the waveguide primarily enters the TE$_{10}$ mode, and only the Stokes sideband is generated through resonant enhancement from the TE$_{00}$ optical mode. 
Based on the fitting of the experimental data, we obtained a -3 dB bandwidth (full width half maximum) of $\sim$1.14 GHz for this modulation effect, which is determined by the optical resonance linewidths. 
We can also quantify the intermodal optomechanical coupling coefficient as $\beta=0.209$ GHz/nm (details in Supplement \S S4).

On the other hand, when light enters the system in the forward direction where the momentum matching condition is not satisfied, very small light scattering is observed (Fig.~\ref{fig5:is}a,iii). 
At the laser detuning where the maximum sideband amplitude is obtained in the backward direction, the sideband generated for forward probing is $\sim$15 dB smaller than that obtained for a backward laser probe. The measurement presented in Fig.~\ref{fig5:is}a,iii is magnified by 10x in order for the data trends to be observable.
The residual scattering that is observed has the characteristic functional shape of conventional intramodal modulation occurring from optical path length change. 
While this effect should nominally be zero, there are practical constraints associated with non-zero overlap integral due to acoustic-optical misalignment and the curvature of the racetrack.

Next we examine the case of mode pair 2, where the optical modes have a larger (8.01 GHz) frequency separation but have identical momentum relationship (Fig.~\ref{fig5:is}b). Here, the modes are frequency-mismatched with respect to the acoustic pump frequency, resulting in much lower indirect intermodal scattering.
For a backward optical probe entering this system, we observe two Lorentzian shapes in both Stokes and anti-Stokes sideband data, corresponding to some intermodal conversion even though the phonon stimulus is non-resonant. 
The larger Lorentzian signature appears due to scattering from the peak of the TE$_{00}$ resonance to off-resonance on the TE$_{10}$ mode. Conversely, the smaller Lorentzian signature corresponds to scattering from an off-resonance point on the TE$_{00}$ mode to the peak of the TE$_{10}$ mode.

In the case of mode pair 3, opposite to mode pairs 1 and 2, the frequency of the TE$_{00}$ mode is lower than the TE$_{10}$ mode by 11.76 GHz.
The phase matching in this case is thus only possible to satisfy in the opposite direction, i.e. for forward optical probing, corresponding to the situation shown in Fig.~\ref{fig1:PM}c.
Measurements of light scattering for this mode pair (Fig.~\ref{fig5:is}c) clearly show greater intermodal conversion for forward input light, even though the frequency matching for the modes is poor. Backward optical probing exhibits much smaller sidebands due to the momentum mismatch.
All the above measurements clearly showcase how the 2-dimensional acoustic pump can be used to satisfy frequency and momentum phase-matching in either forward or backward directions, while also producing the necessary transverse overlap integral.

\vspace{12pt}

\begin{figure}[!t]
	\begin{adjustwidth}{-1in}{-1in}
		\makebox[\textwidth][c]{\includegraphics[width=0.6\textwidth]{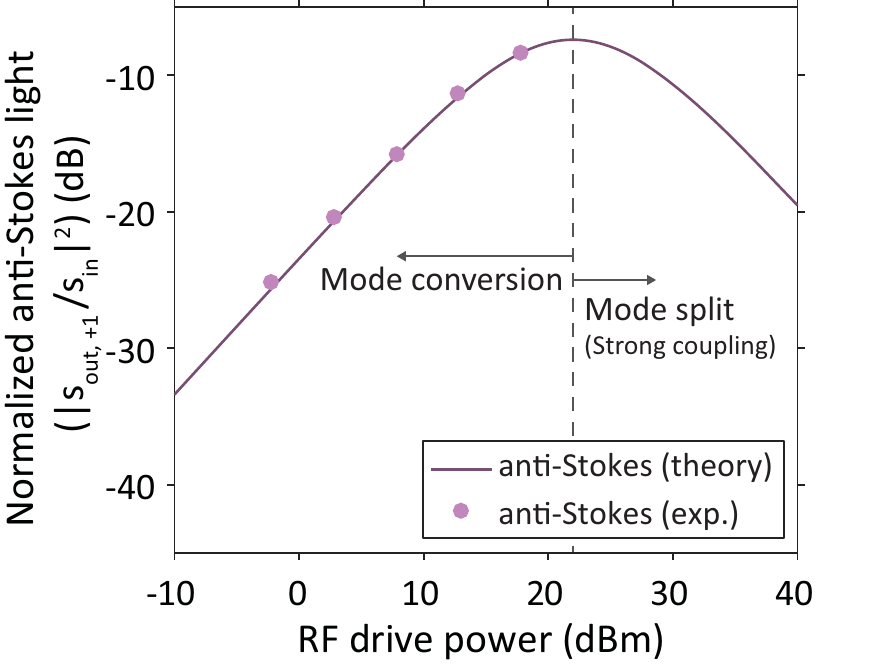}}
		\centering
		\caption{Modulation efficiency measured as a function of RF drive power for mode pair 1 (Fig.~5a). The probe laser is fixed on the resonance of TE$_{00}$ mode in the backward direction (phase matched direction). The theoretical curve (solid line) is calculated based on fitting the data from Fig. 5a, and predicts that the system enters the strong coupling regime when RF drive power goes beyond 21 dBm. Beyond this point the coupled optical modes exhibit splitting and sideband conversion reduces, essentially turning the device into a linear optical isolator}
		\label{fig6:eff}
	\end{adjustwidth}
\end{figure}

In order to test the maximum mode conversion efficiency available with strong acoustic pumping, we perform an experiment using the phase-matched mode pair 1 (Fig.~\ref{fig5:is}a).
The probe laser is applied in the backward, i.e. phase-matched, direction and is detuned from the TE$_{00}$ resonance at the offset where the maximum anti-Stokes sideband is generated. 
The RF power stimulus to the IDT is then swept from -2.8 dBm to 17.8 dBm while the anti-Stokes sideband strength is measured (Fig.~\ref{fig6:eff}).
The solid line in Fig.~\ref{fig6:eff} is a theoretical curve produced using experimentally measured parameters (mode linewidths, coupling rates, optomechanical coupling coefficient) from fitting the data previously shown in Fig.~\ref{fig5:is}a. 
At low drive power, when the sideband amplitude is much smaller than the carrier intracavity field, the mode conversion efficiency linearly increases with the pump. 
In our experiment, we were able to achieve a maximum of 17\% sideband conversion efficiency on resonance  when using 17.8~dBm RF input power.
We calculate that the optomechanical coupling rate at this drive power is $G_{ph} = 0.609$ GHz. 
In a resonant structure, maximum sideband conversion is achievable at the equilibrium point where the amplitudes of the sideband and intracavity optical carrier field are matched (see Supplement \S S3) since the rate of up and down conversion are then equal.
Pumping beyond this point, i.e. $\lvert G _{ph}\rvert^2 > \kappa_1\kappa_2 / 4$, pushes the system into the strong coupling regime where the coupled optical modes begin to split (discussion in Supplement \S S3).
In this regime, for even stronger pumping, the sideband amplitude decreases while the optical carrier frequency component propagates nonreciprocally. In other words, the system begins to operate as a linear optical isolator. 
Unfortunately, due to IDT power limitations we were unable to reach the strong coupling regime in this experiment. However, partial verification of access to this regime is observable in Fig.~\ref{fig6:eff} since the experimental results follow precisely the curvature of the predicted relationship. 
We again emphasize that the solid line in Fig.~\ref{fig6:eff} is not a curve fit, but is a prediction of sideband field relative to the RF input. The conversion efficiency of this system could be improved tremendously by forming an acoustic waveguide (i.e. transverse acoustic resonator) by using free edge reflectors on both sides of the optical waveguide as demonstrated in~\cite{Shin2013}.

\vspace{12pt}

The nanophotonic system that we have presented operates as a frequency shifting isolator in which light propagating in one direction experiences a fixed frequency offset, while in the opposite direction light is simply absorbed.
This type of non-reciprocal device can play an important role in compact atomic timekeeping~\cite{Knappe:04,Esnault:13}, cold-atom inertial navigation~\cite{Gustavson1997}, and gravimetry~\cite{Peters:01} in which magnet-induced Zeeman shifts and light shifts ~\cite{Blanshan:05} are extremely undesirable.
Moreover, the operational optical wavelength and pumping strategy of this system are entirely lithographically defined, which ensures rapid adaptability to other wavelength regimes. 
%, e.g. 780 nm operation for probing Rb, and 854 nm operation for probing Cs.
%
More broadly, this acoustic pumping technique and the level of performance achieved indicates a clear path towards foundry-compatible integration of linear isolators, circulators, and non-reciprocal phase shifters, that overcome the fundamental challenges currently plaguing state of the art non-magnetic non-reciprocal devices. 
This approach can also potentially introduce new non-reciprocal functionality to chip-scale photonics including frequency shifters \cite{Fan2016} and dynamic converters for mode division multiplexing \cite{Luo2014}.

\vspace{24pt}

\FloatBarrier

% ==================================
% ==================================

\section*{Methods}

\subsection*{Device fabrication}

\noindent
We directly deposited c-axis oriented 350 nm film of aluminum nitride (AlN) by RF sputtering onto a silicon wafer. The AlN layer quality was confirmed through X-ray rocking curve measurement and stress measurement. 
The measured full width at half maximum from the rocking curve was 1.78 degree. The average stress of the AlN film was measured as -3.8 MPa (in compression). 
The devices were patterned through three electron-beam lithography steps. 
First, the AlN waveguide and racetrack resonator were patterned using e-beam lithography (Raith E-line) on ZEP-520 resist (ZEON corporation), followed by etching of 200 nm AlN using Cl$_2$ based inductively coupled plasma reactive ion etching (ICP-RIE). Next, release holes and the acoustic edge reflector were patterned using e-beam lithography on double spin coated ZEP-520 resist, followed by complete etch back of the 350 nm AlN through ICP-RIE. 
Finally, the interdigitated transducers (IDTs) were patterned using e-beam lithography on PMMA photoresist after which 60 nm of Al was deposited using an e-beam evaporator. A subsequent lift-off process defines the aluminum IDTs. 
Finally, a gas-phase isotropic silicon etch was performed using XeF$_2$ to release the device.

\subsection*{Mechanical transduction and electronic characterization}

The IDT is used for exciting Lamb wave acoustic modes on the AlN piezoelectric substrate. 
RF signals are provided to the IDT via a ground-signal-ground (GSG) probe (Cascade Microtech model ACP 40). 
The transduction efficiency of the IDT was characterized using the standard approach using a vector network analyzer (Keysight model E5063A) through measurement of the reflection coefficient (s$_{11}$). Details on this measurement are provided in the Supplement \S S4.
The vector network analyzer was calibrated using an on-chip impedance calibration standard to remove any effect of cables and the GSG probe.

\subsection*{Optical measurements}

For performing measurements of intermodal light scattering, the experimental setup shown in Fig.~\ref{fig4:mea}a was used. 
An optical switch (Thorlab model OSW22-1310E) was used to control the light propagation direction, either forward or backward through the optical waveguide.
We used a 1520-1570 nm tunable external cavity diode laser (New Focus model TLB-6728-P, $<$50 kHz instantaneous linewidth) to generate the optical probe. 
The laser source is split 50:50 into a device path and a reference path for performing heterodyne detection of the scattered light spectrum (details in Supplement \S S2). 
Light in the device path was coupled to the on-chip waveguide through grating couplers. An erbium-doped fiber amplifier (EDFA) was used to amplify the light exiting the waveguide to facilitate detection. 
Light in the reference arm was provided a predetermined offset (100 MHz) using an in-fiber acousto-optic frequency shifter (AOFS, Brimrose model AMF-100-1550). 
The beat note between the reference light and the scattered light was measured by the high frequency photodetector (Newport model 1554 photoreceiver). 

\section*{Acknowledgements}

This material is based on research sponsored by Air Force Research Laboratory (AFRL) under agreement number FA9453-16-1-0025. The U.S. Government is authorized to reproduce and distribute reprints for Governmental purposes notwithstanding any copyright notation thereon. The views and conclusions contained herein are those of the authors and should not be interpreted as necessarily representing the official policies or endorsements, either expressed or implied, of Air Force Research Laboratory (AFRL) and (DARPA) or the U.S. Government.

\newpage
\bibliographystyle{myIEEEtran}
\bibliography{main_Ref}

\end{document}

\title{Supplementary information:\\
	 Breaking time-reversal symmetry with
	 acoustic pumping of nanophotonic circuits}

\author{Donggyu Sohn, Seunghwi Kim, Gaurav Bahl$^*$, \\
	\\
	\footnotesize{Mechanical Science and Engineering, University of Illinois at Urbana-Champaign}\\
	\footnotesize{Urbana, Illinois 61801, USA}\\
	\footnotesize{$^\ast$ To whom correspondence should be addressed; }\\
	\footnotesize{E-mail: bahl@illinois.edu}
}

	\maketitle
%	\date{}
%\vspace{24pt}
%\begin{quote} \bf	
%This PDF file includes:
%\end{quote}	
%
%\begin{quote}
%Materials and Methods \\
%SOM Text\\
%Figs. S1 to S3\\
%Table S1\\
%References
%\end{quote}

	\tableofcontents
	
	\newpage

		\newpage

%%%%%%%%%%%%%%%%%%%%%%%%%%%%%%%%
\section{System Model } % rename the title as consistent to your paper
%%%%%%%%%%%%%%%%%%%%%%%%%%%%%%%%
%
% You need to describe the brief introduction which includes transition.

\subsection{Dynamics of Traveling Acousto-optical Interaction}		

Our system consists of two optical modes in a racetrack resonator (quasi-TE$_{10}$ ($\omega_1$, $k_1$) and quasi-TE$_{00}$ ($\omega_2$, $k_2$)) that are coupled by means of an acoustic pump ($\Omega$, $q$) (Fig. \ref{fig:s1}, \ref{fig:s2}). The directionality of the acoustic pump is pre-selected via the electromechanical driving stimulus. With respect to this pump, we define "forward" as the direction in which the light and the acoustic pump are co-propagating, and "backward" as the counter-propagating direction.

The interaction Hamiltonian categorized to the forward and backward direction can be expressed as:
%
%\begin{align*}
%\hbar(\hat{b}^{\dagger} + \hat{b})(\beta^{*} \hat{a}_1^{\dagger} \hat{a}_2 + \beta \hat{a}_1 \hat{a}_2^{\dagger}) 
%\end{align*}
%
%
	\begin{align}
	H_{\text{int}} =  \hbar (\beta_f \hat{a}_1 \hat{a}_2^{\dagger} \hat{b} + \beta^*_f \hat{a}_1^{\dagger} \hat{a}_2 \hat{b}^{\dagger}) + \hbar (\beta_b \hat{a}_1 \hat{a}_2^{\dagger} \hat{b}^{\dagger} + \beta^*_b \hat{a}_1^{\dagger} \hat{a}_2 \hat{b})
	\end{align}\label{eq:s1} 
 %$k_1$, $k_2$, and $q$ are the propagation constant of TE$_{10}$, TE$_{00}$ and the acoustic wave, respectively.
 \noindent where $\beta_f=\beta\,\delta(k_1-k_2+q)$ and $\beta_b=\beta\,\delta(-k_1+k_2+q)$ are the optomechanical coupling coefficients including the phase matching conditions in the forward and backward directions respectively. Here, $\delta ()$ is the Kronecker delta function, $\hat{a}_1^{\dagger} (\hat{a}_1)$ and $\hat{a}_2^{\dagger}(\hat{a}_2)$ are the creation (annihilation) operators for the TE$_{10}$ and TE$_{00}$ modes respectively, and $\hat{b}^{\dagger}(\hat{b})$ is for the acoustic pump~\cite{Agrawal2013, bowen2015quantum}. 
 %Note that optomechanical coupling coefficient~$\beta \propto \int d^2r \phi_1 \phi_2 \psi$ where $\phi_1$, $\phi_2$, and $\psi$ are cross-sectional mode shapes of optical and mechanical modes, respectively. 
%\delta(-k_1+k_2+q)  \delta(k_1-k_2+q)
%

In addition, we include the RF-electromechanically driven acoustic field ($s_{RF}$). The input acoustic field is defined as $|s_{RF}|^2 = \eta_{\text{a}} P_{RF}/ \hbar \Omega$ where $P_{RF}$ is the RF driving power, $\Omega$ is the RF driving frequency, $\eta_a$ is the electromechanical coupling, and $\Gamma$ is the mechanical loss rate. An externally provided laser (s$_{in}$) is coupled to the TE$_{10}$ and TE$_{00}$ modes through the external coupling rates $\kappa_{ex1}$ and $\kappa_{ex2}$. As above, the input optical field is defined as $|s_{in}|^2 = P_{in}/ \hbar \omega_l$ where $P_{in}$ is the input laser power and $\omega_l$ is the frequency of carrier laser. Therefore, the total effective Hamiltonian for the system can be written as~\cite{Bochmann2013}:
%We also include the resonator-waveguide couplings, $\kappa_1$ and $\kappa_2$, for the TE$_{10}$ and TE$_{00}$ mode respectively. The externally provided carrier laser (s$_{in}$) acts through these coupling. Similarly, the input optical field is defined as $|s_{in}|^2 = P_{in}/ \hbar \omega_l$ where $P_{in}$ is the input laser power and $\omega_l$ is the frequency of carrier laser. The total effective Hamiltonian becomes~\cite{Bochmann2013}:
% Please revise the above paragraph if you do not like it.
%
%
\begin{align}
H_{\text{eff}} & = \hbar \omega_1 \hat{a}_1^{\dagger}\hat{a}_1 +  \hbar \omega_2 \hat{a}_2^{\dagger} \hat{a}_2 + \hbar \omega_m \hat{b}^{\dagger} \hat{b} \nonumber\\&+  \hbar\, (\beta_f \hat{a}_1 \hat{a}_2^{\dagger} \hat{b} + \beta^*_f \hat{a}_1^{\dagger} \hat{a}_2 \hat{b}^{\dagger}) + \hbar \,(\beta_b \hat{a}_1 \hat{a}_2^{\dagger} \hat{b}^{\dagger} + \beta^*_b \hat{a}_1^{\dagger} \hat{a}_2 \hat{b}) \nonumber \\ 
&+ i \hbar \sqrt{\kappa_{ex1}}\, (\hat{a}_1^{\dagger} s_{in} e^{-i\omega_l t } - \hat{a}_1 s_{in}^{*} e^{i\omega_l t } )
+i \hbar \sqrt{\kappa_{ex2}}\, (\hat{a}_2^{\dagger} s_{in} e^{-i\omega_l t } - \hat{a}_2 s_{in}^{*} e^{i\omega_l t } ) \nonumber\\ 
&+ i\hbar \sqrt{\Gamma} \,(\hat{b}^{\dagger}s_{RF} e^{-i\Omega t} - \hat{b} s_{RF}^* e^{i\Omega t})
\label{eq:s2}
\end{align}
%
%Here $s_{RF}$ is the RF driven electric amplitude defined as $|s_{RF}|^2 = \eta_{\text{a}} P_{\text{in}}/ \hbar \Omega$, where $\Omega$ is the RF driving frequency, $\eta_a$ is electromechanical coupling coefficient, $s_{in}$ is the carrier laser amplitude defined as $|s_{in}|^2 = P_{in}/ \hbar \omega_l$, $P_{in}$ is the input RF power into the system, $\kappa_{ex1}$ ($\kappa_{ex2}$) are the external coupling rates of $a_1$ ($a_2$) modes to the waveguide. 
Working in a frame rotating with the RF driving frequency ($\Omega$), we write the equations of motion for the mechanical and optical modes by means of Heisenberg-Langevin equation:
\begin{subequations}
	\begin{align}
	\dot{\hat{b}}    = &-i(\omega_m - \Omega) \hat{b} - \frac{\Gamma}{2} \hat{b} - i\beta_f^* \hat{a}_1^\dagger \hat{a}_2 e^{i \Omega t} - i\beta_b \hat{a}_1 \hat{a}_2^{\dagger} e^{i \Omega t}+\sqrt{\Gamma}\,s_{RF}\\
	\dot{\hat{a}}_1  = & -i\omega_1\hat{a}_1 - \frac{\kappa_1}{2} \hat{a}_1 - i\beta_f^* \hat{a}_2 \hat{b}^\dagger e^{i \Omega t} - i\beta_b^* \hat{a}_2 \hat{b} e^{-i \Omega t} +\sqrt{\kappa_{ex1}}\,s_{in}e^{-i\omega_{l} t}\\
	\dot{\hat{a}}_2  = & -i\omega_2\hat{a}_2 - \frac{\kappa_2}{2} \hat{a}_2 - i\beta_f \hat{a}_1 \hat{b} e^{- i \Omega t} - i\beta_b \hat{a}_1 \hat{b}^{\dagger} e^{i \Omega t} +\sqrt{\kappa_{ex2}}\,s_{in}e^{-i\omega_{l} t}
	\end{align}\label{eq:s3}
\end{subequations}
\noindent where $\kappa_1$ ($\kappa_2$) are the optical loss rates of the TE$_{10}$ (TE$_{00}$) modes. 

We will now distill the above general equations for the two specific cases in which light enters from the forward and backward directions. 
		
\subsection{Scattering Process in the Forward Direction (Co-propagating Direction)}

%We can treat the equations of motion in Eq.~(\ref{eq:s3}) classically by making substitutions as follows: $\hat{a_i} \rightarrow a_i$ (i = 1, 2) and $ \hat{b} \rightarrow u$. $u$ is the steady state amplitude of the acoustic field under the non-depleted RF pump approximation where $a_1$ ($a_2$) is the amplitudes of TE$_{00}$ (TE$_{10}$) intracavity field.
%In the forward direction, where the following momentum relationship is satisfied ($k_1-k_2+q = 0$), we can drop off the delta function in $\beta_f$ and convert it to $\beta$. Due to the phase matching condition
%Fig.~\ref{fig:s1} shows that due to the phase matching condition ($k_1-k_2+q = 0$),
%As shown in the forward three wave coupling term, TE$_{10}$  can only convert into TE$_{00}$ mode through anti-Stokes scattering. Similarly, the TE$_{00}$ mode can only convert into TE$_{10}$ mode through Stokes scattering. These interaction can be seen in Eq.~(\ref{eq:s4}) below. These equations are obtained by neglecting the non-phase matched terms, i.e. the terms with $\beta_b$ in Eq.~\ref{eq:s3}.

\begin{figure}[t!]
	%	\begin{adjustwidth}{-1in}{-1in}
	\makebox[\textwidth][c]{\includegraphics[width=0.6\textwidth]{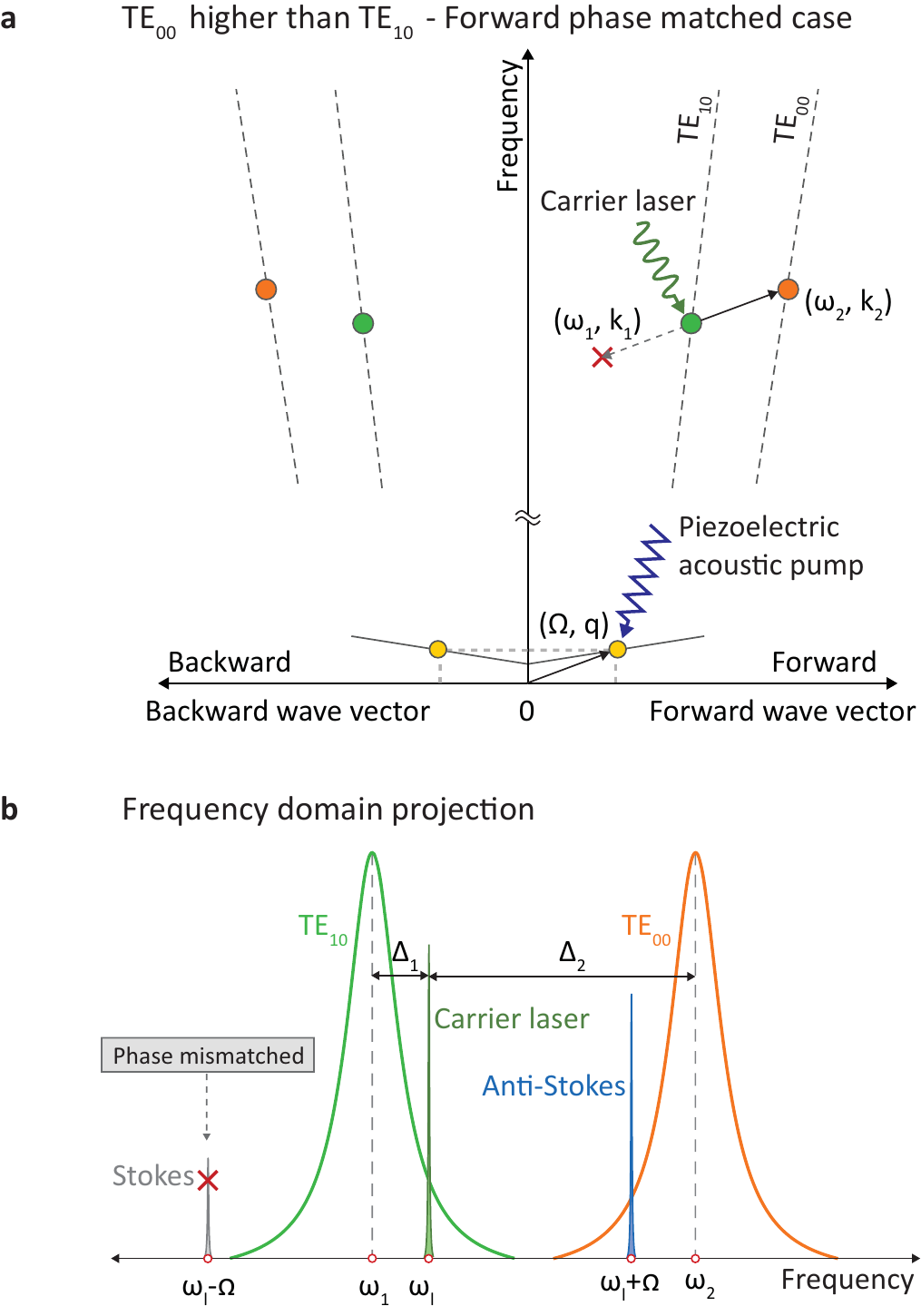}}
	\centering
	\caption{(a) Intermodal scattering illustrated in frequency-momentum space. In the case where the carrier laser is presented to the TE$_{10}$ mode in the forward direction, the anti-Stokes sideband is resonantly enhanced by the TE$_{00}$ resonance mode. The Stokes sideband is suppressed since there is no optical mode available. (b) The intermodal scattering process illustrated in frequency domain. Due to the energy-momentum phase matching condition only the anti-Stokes sideband appears, which is located at the frequency $\omega_l+\Omega$.}
	\label{fig:s1}
	%	\end{adjustwidth}
\end{figure}

We treat the equations of motion in Eq.~(\ref{eq:s3}) classically by making substitutions as follows: $\hat{a}_1, (\hat{a}_2) \rightarrow a_1$ ($a_2$) and $ \hat{b} \rightarrow u$. Thus $a_1$ and $a_2$ are the amplitudes of the TE$_{10}$ and TE$_{00}$ intracavity fields respectively and $u$ is the steady state amplitude of the acoustic field under the non-depleted RF pump approximation. We consider the situation illustrated in Fig.~\ref{fig:s1}, where phase matching only occurs for forward propagating light ($k_1-k_2+q = 0$). Here, the non-phase matched terms can be neglected, i.e. the term with $\beta_b$ in Eq.~\ref{eq:s3}. We can also drop $\delta ()$ in the forward intermodal coupling coefficient and simply use $\beta$ as the optomechanical coupling rate. The TE$_{10}$ mode can only convert into the TE$_{00}$ mode through anti-Stokes scattering. Similarly, the TE$_{00}$ mode can only convert into the TE$_{10}$ mode through Stokes scattering. Thus, we can rewrite Eq.~\ref{eq:s3} as:

%as shown in the coupling term in Eq.~(\ref{eq:s4}). Any coupling term that does not satisfies the phase matching relation is neglected e.g. the term with $\beta_b$. Since , which is of key interest in the paper, we obtain the simplified equations:
%For the analytical description of the intermodal scattering, we used mechanical formalism shown in Agarwal []. 
	%As shown in In the forward direction ($k_1-k_2+q=0$), 
\begin{subequations}
	\begin{align}
	\dot{a_1} 	= & -\frac{\kappa_1}{2} a_1 - i\omega_1 a_1 - i\beta^*u^* a_2 e^{i\Omega t}+\sqrt{\kappa_{ex1}}\,s_{in}e^{- i\omega_l t}\\
	\dot{a_2}	= & -\frac{\kappa_2}{2} a_2 - i\omega_2 a_2 - i\beta u a_1 e^{-i\Omega t}+\sqrt{\kappa_{ex2}}\,s_{in}e^{- i\omega_l t}
	\end{align}
	\label{eq:s4}
\end{subequations}
   %$\omega_1$ and $\omega_2$ are optical resonance frequency of TE$_{00}$ mode and TE$_{10}$ mode respectively,
%$\omega_l$ is the carrier laser frequency, $\Omega$ is the applied mechanical frequency, $\beta$ is the intermodal optomechanical coupling coefficient that account modal overlap integral of three waves.

The mode amplitudes $a_1$ and $a_2$ can be now expressed through Fourier decomposition of the sidebands created by the intermodal scattering:
\begin{subequations}
	\begin{align}
	a_1 	= &\sum_{n=-\infty}^{\infty} a_{1,n} e^{- i(\omega_l + n\Omega)t} ~=~a_{1,-1}e^{ -i (\omega_l-\Omega)t} + a_{1,0}e^{ - i\omega_lt}\\
	a_2		= &\sum_{n=-\infty}^{\infty} a_{2,n} e^{- i(\omega_l + n\Omega)t} ~=~a_{2,0} e^{ - i\omega_lt}+a_{2,+1}e^{ - i(\omega_l + \Omega)t}
	\end{align}
	\label{eq:5}	
\end{subequations}

\noindent where n indicates the sideband order. Here, we adopted the following convention: $a_{i,0}$ is the intracavity field at the carrier frequency, while $a_{i,-1}$ ($a_{i,+1}$) are the intracavity fields at the Stokes (anti-Stokes) shifted frequency for the $i$th order mode. Thus, $a_{1,0}$ ($a_{2,0}$) are the intracavity field amplitude for TE$_{10}$ (TE$_{00}$) mode at the carrier frequency i.e. the frequency of the source laser. $a_{1,-1}$ is the Stokes sideband amplitude for the TE$_{10}$ mode. $a_{2,+1}$ is the anti-Stokes sideband amplitude for the TE$_{00}$ mode. The remaining sidebands are suppressed due to the phase matching condition, i.e. absence of suitable optical modes. 

The intracavity field amplitudes $a_{1,0}$, $a_{2,0}$, $a_{1,-1}$ and $a_{2,+1}$ are obtained by substituting Eq.~(\ref{eq:5}) to Eq.~(\ref{eq:s4}).
\begin{align}
	a_{1,0}	 &=\frac{\sqrt{\kappa_{ex1}}\,s_{in}-i\beta^*u^*a_{2,+1}}{\kappa_1 /2 - i\Delta_1} &\qquad\text{and}\qquad&
	a_{2,0}	 = \frac{\sqrt{\kappa_{ex2}}\,s_{in}-i\beta u a_{1,-1}}{\kappa_2 /2 - i\Delta_2} \nonumber\\
	a_{1,-1} &=\frac{-i\beta^* u^* a_{2,0}}{\kappa_1 /2 - i(\Delta_1-\Omega)} 
	&\qquad\text{and}\qquad&
	a_{2,+1} = \frac{-i\beta u a_{1,0}}{\kappa_2 /2 - i(\Delta_2+\Omega)}
		\label{eq:6}
\end{align}
where the optical detunings are $\Delta_1 = \omega_l - \omega_1$ and $\Delta_2 = \omega_l - \omega_2$.

In the solutions for the carrier frequency components ($a_{1,0}$ and $a_{2,0}$), the first term in the numerator is the input field coupled from the optical waveguide and the second term is the field scattered back from the corresponding sideband. The sideband intracavity fields can be finally expressed in terms of the input laser $(s_{in})$ as follows:

\begin{subequations}
	\begin{align}
	a_{1,0} &= \frac{\sqrt{\kappa_{ex1}} \left( \kappa_2/2 - i(\Delta_2+\Omega) \right) }{\left(\kappa_2/2 - i(\Delta_2+\Omega)\right)\left(\kappa_1/2 - i\Delta_1\right)+\lvert \beta \rvert^2 \lvert u \rvert^2} s_{in}, \\
	a_{2,0} &= \frac{\sqrt{\kappa_{ex2}}(\kappa_1 /2 - i(\Delta_1-\Omega))}{(\kappa_1/ 2 - i(\Delta_1-\Omega))(\kappa_2/2 - i\Delta_2)+\lvert \beta \rvert^2 \lvert u \rvert^2} s_{in}, \\
	a_{1,-1} &=\frac{-i\beta^* u^*\sqrt{\kappa_{ex2}} }{(\kappa_1 /2  - i(\Delta_1-\Omega))(\kappa_2 /2 - i\Delta_2)+\lvert \beta \rvert^2 \lvert u \rvert^2} s_{in},  \\
	a_{2,+1} &=\frac{-i\beta u\sqrt{\kappa_{ex1}}}{(\kappa_2 /2 - i(\Delta_2+\Omega))(\kappa_1 /2 - i\Delta_1)+\lvert \beta \rvert^2 \lvert u \rvert^2} s_{in}.
	\end{align} \label{eq:7}
\end{subequations}

%\begin{subequations}
%\begin{align}
%a_{1,0} &= \frac{\sqrt{\kappa_{ex1}} \left( \frac{\kappa_2}{2} - i(\Delta_2+\Omega) \right) }{\left(\frac{\kappa_2}{2} - i(\Delta_2+\Omega)\right)\left(\frac{\kappa_1}{2} - i\Delta_1\right)+\lvert \beta \rvert^2 \lvert u \rvert^2} s_{in}, \\
%a_{2,0} &= \frac{\sqrt{\kappa_{ex2}}\left( \frac{\kappa_1}{2} - i(\Delta_1-\Omega) \right)}{ \left( \frac{\kappa_1}{2} - i(\Delta_1-\Omega)\right) \left( \frac{\kappa_2}{2} - i\Delta_2 \right) +\lvert \beta \rvert^2 \lvert u \rvert^2} s_{in}, \\
%a_{1,-1} &=\frac{-i\beta^* u^*\sqrt{\kappa_{ex2}} }{( \left( \frac{\kappa_1}{2} - i(\Delta_1-\Omega)\right) \left( \frac{\kappa_2}{2} - i\Delta_2 \right) +\lvert \beta \rvert^2 \lvert u \rvert^2} s_{in},  \\
%a_{2,+1} &=\frac{-i\beta u\sqrt{\kappa_{ex1}}}{\left(\frac{\kappa_2}{2} - i(\Delta_2+\Omega)\right)\left(\frac{\kappa_1}{2} - i\Delta_1\right)+\lvert \beta \rvert^2 \lvert u \rvert^2} s_{in}.
%\end{align} \label{eq:7}
%\end{subequations}	
%where $\Delta_1 = \omega_l - \omega_1$ and $\Delta_2 = \omega_l - \omega_2$.

%
When the optomechanical coupling rate is much smaller than the optical loss rate ($\lvert \beta \rvert \lvert u \rvert << \sqrt{\kappa_1\kappa_2}/2$), e.g. for small acoustic drive ($u$), Eq.~\ref{eq:6} can be simplified to:

\begin{subequations}
	\begin{align}
	a_{1,0} &	=\frac{\sqrt{\kappa_{ex1}}}{\kappa_1 /2  - i\Delta_1} s_{in}, \\
	a_{2,0} &	=\frac{\sqrt{\kappa_{ex2}}}{\kappa_2 /2  - i\Delta_2} s_{in}, \\
	a_{1,-1} &	=\frac{-i\beta^* u^*\sqrt{\kappa_{ex2}}}{(\kappa_1 /2 - i(\Delta_1-\Omega))(\kappa_2 /2 - i\Delta_2)} s_{in}, \\
	a_{2,+1} &	=\frac{-i\beta u\sqrt{\kappa_{ex1}}}{(\kappa_2 /2 - i(\Delta_2+\Omega))(\kappa_1 /2 -i\Delta_1)} s_{in}
	\end{align}
\end{subequations}	

By means of input output theorem, we obtain the expression for the output spectrum ($s_{out}$) from the waveguide containing carrier ($s_{out,0}$), Stokes ($s_{out,-1}$), and anti-Stokes ($s_{out,+1}$) frequency components:
\begin{align}
s_{out}= s_{out,0}+ s_{out,+1}e^{-i\Omega t}+ s_{out,-1}e^{i\Omega t}
\label{eq:9}
\end{align}
where
\begin{subequations}
	\begin{align}
	s_{out,0} &	=s_{in}-\sqrt{\kappa_{ex1}}\,a_{1,0}-\sqrt{\kappa_{ex2}}\,a_{2,0}\\
	s_{out,-1} & =-\sqrt{\kappa_{ex1}}\,a_{1,-1}  	\label{eq:s10b}\\
	s_{out,+1} & =-\sqrt{\kappa_{ex2}}\,a_{2,+1}
	\end{align}
\end{subequations}	

These equations are used to plot the theoretical curves in Fig.~5 of the main manuscript. 
\subsection{Scattering Process in the Backward Direction (Counter-propagating Direction)}

\begin{figure}[b!]
	%	\begin{adjustwidth}{-1in}{-1in}
	\makebox[\textwidth][c]{\includegraphics[width=0.65\textwidth]{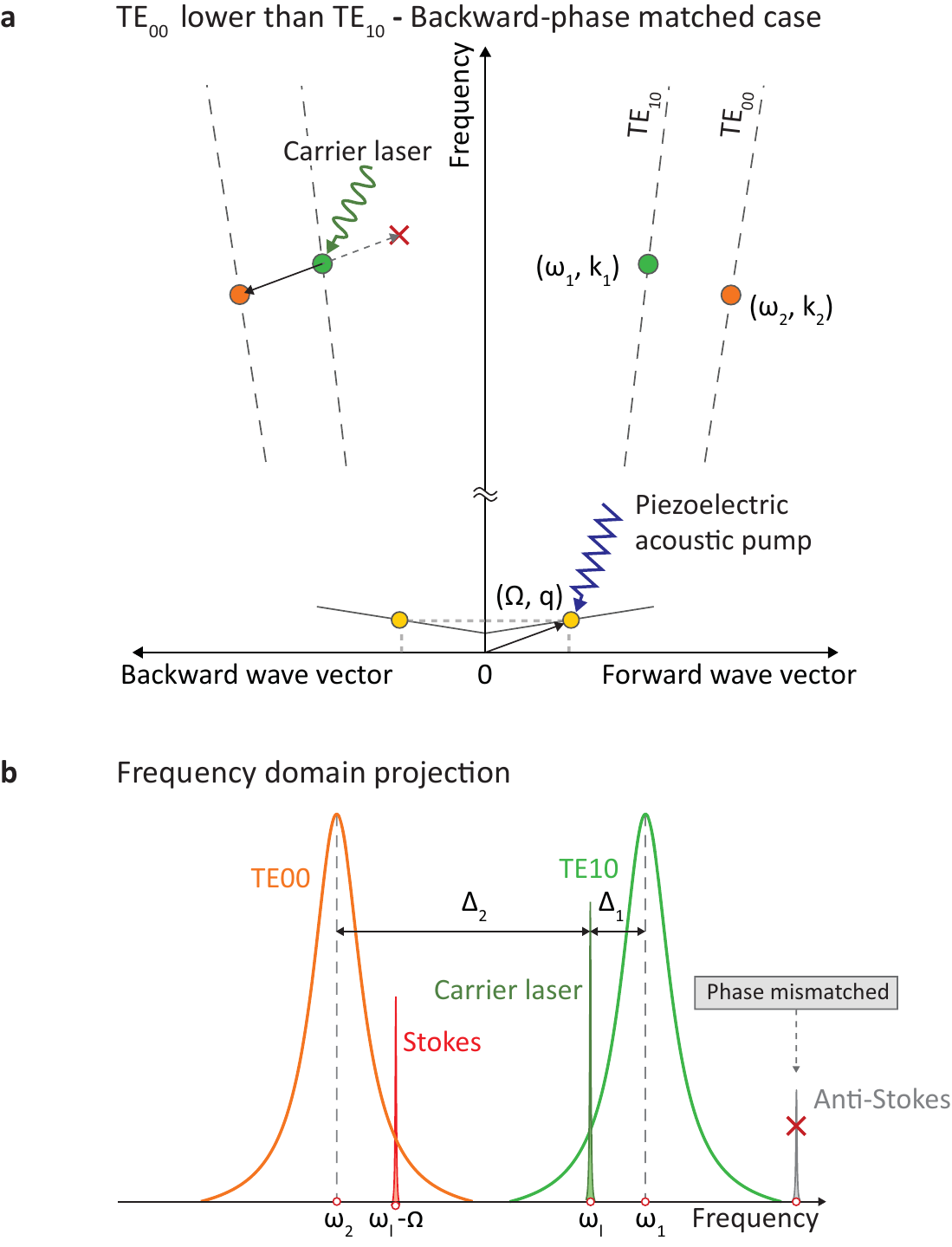}}
	\centering
	\caption{(a) Intermodal scattering illustrated in frequency-momentum space. In the case where the carrier laser is presented to the TE$_{10}$ mode in the backward direction, the Stokes sideband is resonantly enhanced by the TE$_{10}$ mode. The anti-Stokes sideband is suppressed since there is no optical mode that allows scattering. (b) The intermodal scattering illustrated in frequency domain. Due to the phase matching condition only the Stokes sideband appears, which is located at the frequency $\omega_l-\Omega$.}
	\label{fig:s2}
	%	\end{adjustwidth}
\end{figure}

Let us now consider the case shown in Fig.~\ref{fig:s2} where the light propagating direction is in the backward direction. Due to the backward phase matching condition ($-k_1+k_2+q = 0$), as opposed to the scattering in the forward direction, the TE$_{10}$ (TE$_{00}$) modes can only convert into the TE$_{00}$ (TE$_{10}$) modes through Stokes (anti-Stokes) scattering. When the resonance frequency of the TE$_{10}$ mode is higher than the TE$_{00}$ mode, the phase matching condition is satisfied in the backward direction. Thus, in this case, the equations of motion in Eq.~\ref{eq:s3} are simplified to:
%%When the resonance of the TE$_{00}$ is located in the higher frequency, the light scattered from the resonance immediately exits the resonator since the frequency matching condition is not satisfied (Fig.~\ref{fig:s2}). In this case, amplitude of the sideband is relatively small. However, In the case shown in Fig. 1b (main man
\begin{subequations}
	\begin{align}
	\dot{a_1} 	= & -\frac{\kappa_1}{2} a_1 - i\omega_1 a_1 - i\beta^*u a_2 e^{-i\Omega t}+\sqrt{\kappa_{ex1}}\,s_{in}e^{- i\omega_l t}\\
	\dot{a_2}	= & -\frac{\kappa_2}{2} a_2 - i\omega_2 a_2 - i\beta u^* a_1 e^{i\Omega t}+\sqrt{\kappa_{ex2}}\,s_{in}e^{- i\omega_l t}
	\end{align}
	\label{eq:4}
\end{subequations}
%The intermodal scattering exists when the light enters in the backward direction. Let's consider the case shown in Fig.1a that the acoustic wave is propagating in the backward direction. When the carrier light is presented to TE$_{10}$ resonance in the backward direction, the intermodal scattering can only occur through Stokes scattering to TE$_{00}$ mode. The scattered photons have very large frequency detuning from the resonance so that the photons exit the resonator immediately. We can simply model the scattering in the non-phase matching direction by exchange Eq.~\ref{eq:5} by Eq.~\ref{eq:13}. Here, only Stokes ($a_{2,-1}$) from TE$_{10}$ to TE$_{00}$ and anti-Stokes ($a_{1,+1}$)from TE$_{00}$ to TE$_{10}$ exist. All the other scattering is suppressed due to the phase matching condition.
%When the light enters in the backward direction, the light in the resonance is scattered to the other optical mode that does not satisfy the resonant condition. Therefore, in the frequency domain (Fig.~\ref{fig:s2}), only the Stokes sideband with TE$_{00}$ mode is created. However, since this sideband is located outside of the resonance, amplitude of the sideband is very small. The intermodal scattering in the non-phase matching direction can be observed in the experimental data. 
The composited amplitudes $a_1$ and $a_2$ can be written as: 
\begin{subequations}
	\begin{align}
	a_1 	= &a_{1,0}e^{ -i \omega_l t} + a_{1,+1}e^{ - i(\omega_l+\Omega)t}\\
	a_2		= &a_{2,-1} e^{ - i(\omega_l-\Omega)t}+a_{2,-1}e^{ - i\omega_l t}
	\end{align}
	\label{eq:13}	
\end{subequations}
where $a_{1,+1}$ is the anti-Stokes sideband amplitude in the TE$_{10}$ mode, and $a_{2,-1}$ is the Stokes sideband amplitude in the TE$_{00}$ mode.
\begin{align}
a_{1,0}	 &=\frac{\sqrt{\kappa_{ex1}}\,s_{in}-i\beta^*ua_{2,-1}}{\kappa_1 /2 - i\Delta_1} &\qquad\text{and}\qquad&
a_{2,0}	 = \frac{\sqrt{\kappa_{ex2}}\,s_{in}-i\beta u^* a_{1,+1}}{\kappa_2 /2 - i\Delta_2} \nonumber\\
a_{1,+1} &=\frac{-i\beta^* u a_{2,0}}{\kappa_1 /2 - i(\Delta_1+\Omega)} 
&\qquad\text{and}\qquad&
a_{2,-1} = \frac{-i\beta u^* a_{1,0}}{\kappa_2 /2 - i(\Delta_2-\Omega)}
\label{eq:s13}
\end{align}

The sideband intracavity fields can be expressed in terms of the input laser $(s_{in})$ as follows:  

\begin{subequations}
	\begin{align}
	a_{1,0} &= \frac{\sqrt{\kappa_{ex1}} \left( \kappa_2/2 - i(\Delta_2-\Omega) \right) }{\left(\kappa_2/2 - i(\Delta_2-\Omega)\right)\left(\kappa_1/2 - i\Delta_1\right)+\lvert \beta \rvert^2 \lvert u \rvert^2} s_{in}, \\
	a_{2,0} &= \frac{\sqrt{\kappa_{ex2}}(\kappa_1 /2 - i(\Delta_1+\Omega))}{(\kappa_1/ 2 - i(\Delta_1+\Omega))(\kappa_2/2 - i\Delta_2)+\lvert \beta \rvert^2 \lvert u \rvert^2} s_{in}, \\
	a_{1,+1} &=\frac{-i\beta^* u\sqrt{\kappa_{ex2}} }{(\kappa_1 /2  - i(\Delta_1+\Omega))(\kappa_2 /2 - i\Delta_2)+\lvert \beta \rvert^2 \lvert u \rvert^2} s_{in},  \\
	a_{2,-1} &=\frac{-i\beta u^*\sqrt{\kappa_{ex1}}}{(\kappa_2 /2 - i(\Delta_2-\Omega))(\kappa_1 /2 - i\Delta_1)+\lvert \beta \rvert^2 \lvert u \rvert^2} s_{in}.
	\end{align}
\end{subequations}	

When the optomechanical coupling rate is much smaller than the optical loss rate ($\lvert \beta \rvert \lvert u \rvert << \sqrt{\kappa_1\kappa_2}/2$), e.g. for small acoustic drive ($u$), Eq.~\ref{eq:s13} can be simplified to:

\begin{subequations}
	\begin{align}
	a_{1,0} &	=\frac{\sqrt{\kappa_{ex1}}}{\kappa_1 /2  - i\Delta_1} s_{in}, \\
	a_{2,0} &	=\frac{\sqrt{\kappa_{ex2}}}{\kappa_2 /2  - i\Delta_2} s_{in}, \\
	a_{1,+1} &	=\frac{-i\beta^* u\sqrt{\kappa_{ex2}}}{(\kappa_1 /2 - i(\Delta_1+\Omega))(\kappa_2 /2 - i\Delta_2)} s_{in}, \\
	a_{2,-1} &	=\frac{-i\beta u^*\sqrt{\kappa_{ex1}}}{(\kappa_2 /2 - i(\Delta_2-\Omega))(\kappa_1 /2 -i\Delta_1)} s_{in}
	\end{align} \label{eq:8}
\end{subequations}	

Similar to the output from the previous section, the output spectrum ($s_{out}$) from the waveguide contains the carrier ($s_{out,0}$), Stokes ($s_{out,-1}$), and anti-Stokes ($s_{out,+1}$) frequency components:
\begin{align}
s_{out}= s_{out,0}+ s_{out,+1}\,e^{-i\Omega t}+ s_{out,-1}\,e^{i\Omega t}
	\label{eq:9}
\end{align}
where
\begin{subequations}
	\begin{align}
	s_{out,0} &	=s_{in}-\sqrt{\kappa_{ex1}}\,a_{1,0}-\sqrt{\kappa_{ex2}}\,a_{2,0}\\
	s_{out,-1} & =-\sqrt{\kappa_{ex2}}\,a_{2,-1}\\
	s_{out,+1} & =-\sqrt{\kappa_{ex1}}\,a_{1,+1}
	\label{eq:s17}
	\end{align}
\end{subequations}

Again, these equations are used to plot the theoretical curves in Fig. 5 of the main manuscript. 

%These equations are used to plot the theoreical curves in the main manuscript

%In case when the resonant frequency of TE$_{10}$ mode becomes higher than that of TE$_{10}$, which is illustrated in Figure 1b, the phase matching (both momentum and frequency matching) condition is satisfied in the backward direction. The carrier present in TE$_{10}$ mode is scatter into the resonance of the TE$_{00}$ mode so that the sideband amplitude can be amplified by the second resonance.

\section{Measurement of Optical Spectra}
		
	\begin{figure}[t!]
		%	\begin{adjustwidth}{-1in}{-1in}
		\makebox[\textwidth][c]{\includegraphics[width=0.9\textwidth]{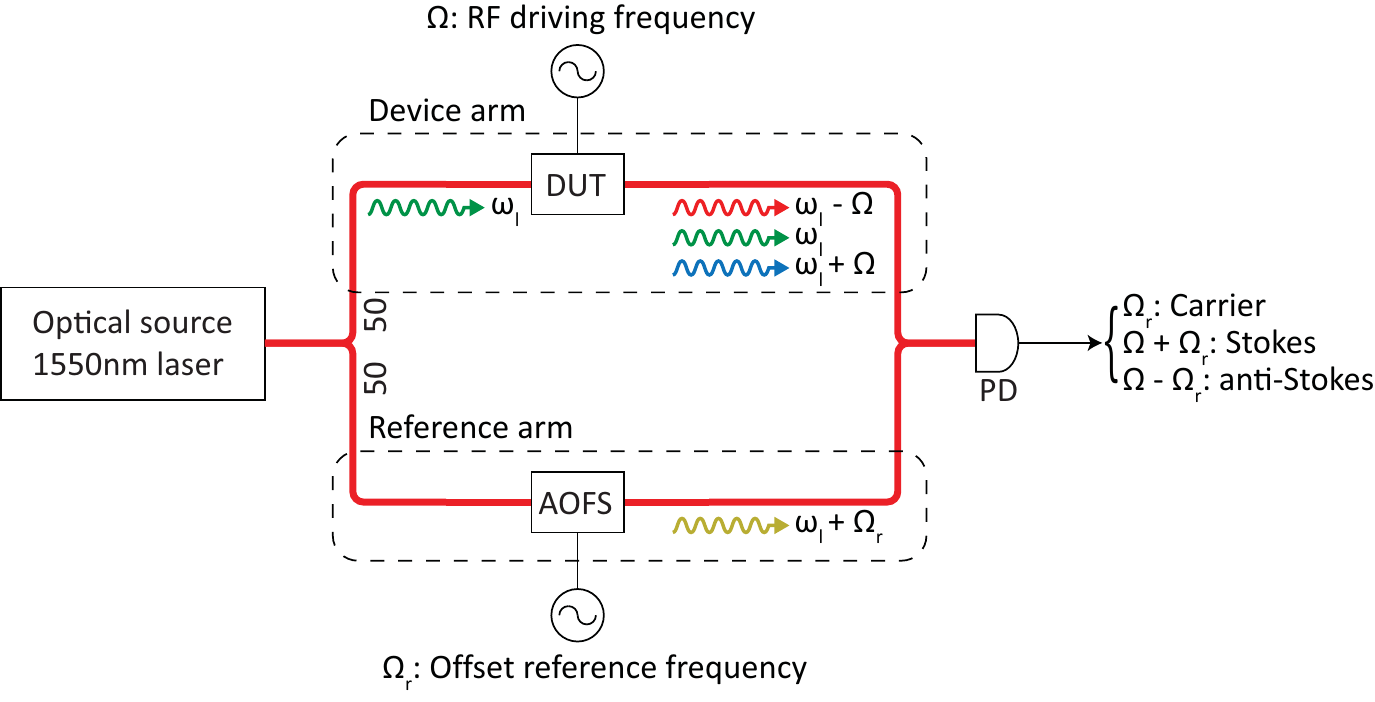}}
		\centering
		\caption{Measurement setup utilizing optical heterodyne detection. The light from the optical source is split into the reference arm and the device arm. The frequency of the light in the reference arm is shifted by the AOFS by predetermined offset $\Omega_r$. The beat notes of the frequency shifted reference with the carrier, Stokes, and anti-Stokes from the device under test (DUT) are measured by the photodetector and have frequency components $\Omega_r$, $\Omega+\Omega_r$, and $\Omega-\Omega_r$ respectively. }
		\label{fig:meas}
		%	\end{adjustwidth}
	\end{figure}	
		
		 In order to measure the amplitude of the Stokes and anti-Stokes sidebands separately, we utilize optical heterodyne detection with assistance of an acousto-optic frequency shifter (AOFS). The detailed experimental setup is shown in Fig.\ref{fig:meas}. 
		 
		 Considering the frequency up-shifted reference signal ($s_r e^{-i\Omega_r t}$) generated from the AOFS in the reference arm, the optical output spectrum at the photodector can be expressed as: 
		 
			\begin{align}
			s_{out}= s_{out,0}+ s_{out,+1}\, e^{-i\Omega t}+ s_{out,-1}\,e^{i\Omega t} + s_r \, e^{-i\Omega_r t}
			\label{eq:10}
			\end{align}
	
	The resulting electronic signals from the photodetector are beat notes of the optical reference signal with the carrier, Stokes, and anti-Stokes signals, and occur at $\Omega_r$, $\Omega+\Omega_r$, and $\Omega-\Omega_r$ respectively. The powers of each frequency component can be independently measured using an electronic spectrum analyzer. Solving Eq.~(\ref{eq:10}) for each frequency component while considering the photodetector gain, the RF outputs are expressed as:
		%Therefore, the optical power from the output waveguide can be expressed as following. 
		%\begin{align}
		%\lvert s_{out} \rvert^2=\lvert-s_{in}+\sqrt{\kappa_{ex1}}(a_{1,0}+a_{1,-1}e^{-i\Omega t})+\sqrt{\kappa_{ex2}}(a_{2,0}+a_{2,+1}e^{i\Omega t})+s_r e^{i\Omega_r t}\rvert^2
		%\end{align}
		%In Eq.S20, the term with $\Omega_r$ can be expressed as following, which is the power 
		%\begin{align}
		%\lvert s_{out,\Omega_r} \rvert^2= 2Re[s_r^* e^{-i\Omega_r t}(-s_{in}+\sqrt{\kappa_{ex1}}a_{1,0}+\sqrt{\kappa_{ex2}}a_{2,0})]=2\lvert s_r %\rvert \lvert -s_{in}+\sqrt{\kappa_{ex1}}a_{1,0}+\sqrt{\kappa_{ex2}}a_{2,0} \rvert cos(\Omega_r t +\theta_\Omega)
		%\end{align}
		%$\Omega-\Omega_r$ term (Antistokes)		
		%\begin{align}
		%\lvert s_{out,\Omega-\Omega_r} \rvert^2= 2Re[s_r^* e^{-i\Omega_r t}(\sqrt{\kappa_{ex2}}a_{2,+1}e^{i\Omega t})]= 2\lvert s_r \rvert \lvert %\sqrt{\kappa_{ex2}}a_{2,+1} \rvert cos\{(\Omega-\Omega_r)t +\theta_{\Omega-\Omega_r}\}
		%\end{align}				
		%$\Omega+\Omega_r$ term (Stokes)		
		%\begin{align}
		%\lvert s_{out,\Omega+\Omega_r} \rvert^2= 2Re[s_r^* e^{-i\Omega_r t}(\sqrt{\kappa_{ex1}}a_{1,-1}e^{-i\Omega t})]=2\lvert s_r \rvert \lvert %\sqrt{\kappa_{ex1}}a_{1,-1} \rvert cos\{(\Omega+\Omega_r)t +\theta_{\Omega+\Omega_r}\}
		%\end{align}	
		%As shown above, the beat notes of transmission, stokes, and Anti-stokes exist in different frequency. Therefore, we are able to individually measure the 
		%Photodetector generates current that is proportional to light power.
		\begin{equation}
		\begin{split}
		P_{C,\Omega_r}&=g_{pd}\lvert s_r \rvert^2 \lvert s_{out,0}  \rvert^2\\
		P_{S,\Omega+\Omega_r}&=g_{pd}\lvert s_r \rvert^2 \lvert s_{out,-1} \rvert^2\\
		P_{AS,\Omega-\Omega_r}&=g_{pd}\lvert s_r \rvert^2 \lvert s_{out,+1} \rvert^2
		\end{split} \label{eq:11}
		\end{equation}
		
		\noindent where $P_{C,\Omega_r}$, $P_{S,\Omega+\Omega_r}$, and $P_{AS,\Omega-\Omega_r}$ are the carrier, Stokes, and anti-Stokes RF power outputs from the photodetector, respectively. Here, $g_{pd}$ is the lumped proportionality constant including photodetector gain. 
		
		Since the acousto-optic scattering process is linearly proportional to the input light power, the optical power from the output waveguide should normalize to the input light power:
		\begin{align*}
		P_{in}=g_{pd}\lvert s_r \rvert^2 \lvert s_{in} \rvert^2
		\end{align*}		
		
		Dividing the measured power by the input power, the normalized power coefficients are given by:
		\begin{equation}
		\begin{split}		
			\bar{P}_{C,\Omega_r}&=\left| \frac{s_{out,0}}{s_{in}} \right|^2\\
			\bar{P}_{S,\Omega+\Omega_r}&=\left| \frac{s_{out,-1}}{s_{in}} \right|^2\\
			\bar{P}_{AS,\Omega-\Omega_r}&=\left| \frac{s_{out,+1}}{s_{in}} \right|^2
		\end{split} \label{eq:12}
		\end{equation}

We use these coefficients to quantify the performance of this mode converting modulator.

\section{Theoretical Limit of Sideband Amplitude}

	We now wish to theoretically quantify the maximum sideband amplitude achievable in this system. Let us recall the intracavity Stokes sideband field amplitude (Eq.~\ref{eq:s10b}) for the TE$_{10}$ mode. Here, we introduce the phonon enhanced optomechanical coupling rate $G_{ph}=\beta u$. The Stokes sideband from the waveguide $s_{out,-1}$ is:

	\begin{align}
	s_{out,-1}&=\sqrt{\kappa_{ex1}}\,\left( \frac{-iG_{ph}^*\sqrt{\kappa_{ex2}} }{(\kappa_1 /2  - i(\Delta_1-\Omega))(\kappa_2 /2 - i\Delta_2)+ \lvert G_{ph} \rvert^2}\right) s_{in}
	\end{align}	

\noindent From the above equation, we can see that the sideband amplitude is maximized when the frequency matching between the modes is perfect ($\omega_1-\omega_2+\Omega = 0$) and the carrier is located on the resonance of the TE$_{00}$ mode. As a result, the imaginary parts in the denominator vanish and the equation can be simplified to:
	\begin{align}
	s_{out,-1}= \frac{-iG_{ph}^*\sqrt{\kappa_{ex1}\kappa_{ex2}} }{\kappa_1\kappa_2 /4+ \lvert G_{ph} \rvert^2} s_{in}
	\label{eq:s18}
	\end{align}	
To investigate the thoeretical limit of sideband amplitude with respect to phonon-enhanced optomechanical coupling $G_{ph}$, we calculate the maximum point where $G_{ph}$ satisfies $\frac{\partial s_{out,-1}}{\partial G} = 0$, which is:

\begin{align}
G_{ph} = \frac{\sqrt{\kappa_1\kappa_2}}{2}
\label{eq:s19}
\end{align}	
\noindent At this pump level, the calculated maximum sideband field amplitude becomes:
\begin{equation}
s_{out,-1}\rvert_{max}=\sqrt{\frac{\kappa_{ex1}\kappa_{ex2}}{\kappa_1\kappa_2}} \, s_{in}
\label{eq:s20}
\end{equation}	
This is the point where the photon energy exchange between the two optical modes $a_1$ and $a_2$ is in equilibrium. From the equation, we can see that if there is no intrinsic photon loss in the resonator, which means that the external couplings are only photon loss mechanism ($\kappa_{ex1}=\kappa_1, \kappa_{ex2}=\kappa_2$), 100\% of light can be transfered to the sideband. In case where both optical modes are critically coupled to the waveguide ($\kappa_{ex1}=\kappa_1/2, \kappa_{ex2}=\kappa_2/2$), the maximum sideband amplitude is 25\% of carrier power (50\% of carrier field). The rest of power is dissipated in the resonator. If the acoustic pump increases beyond this point, the optical modes split, i.e the system enters the strong coupling regime (see Supplement \S S5).

%
%	
%	
%	
%	In the equation, when the $\beta \u$
%	
%	There is maxmium theoretically quantify the maximum sideband amplitude 
%
%As shown in Eq.~\ref{eq:8}, the sideband amplitude can be expressed as:

%In an optically resonant structure, where the spatial variation of the optical field is negligible, theoretical limit of sideband amplitude exists. In other words, full mode conversion is not allowed although the sufficient acoustic power is applied to system. Let's consider the sideband field amplitude shown in Eq.~\ref{eq:7}d. Here, for simplicity, we assume that mechanical frequency is the same with the mode separation and the carrier laser is presented on the resonance of $a_1$ mode ($\Delta_1 = 0$). Since the maximum sideband amplitude is achievable at the maxima point ($\lvert \beta \rvert \lvert u \rvert=\sqrt{\kappa_1\kappa_2}/2$) of Eq.~\ref{eq:7}d i.e. $\lvert \beta \rvert \lvert u \rvert$ satisfied with $\frac{\partial \lvert \sqrt{k_{ex2}}a_{2,+1}\rvert}{\partial \lvert \beta \rvert \lvert u \rvert} = 0$, the maximum normalized sideband amplitude from the waveguide is expressed as:
%%
%%The maxima $\frac{\partial \lvert \sqrt{k_{ex2}}a_{2,+1}\rvert}{\partial \lvert \beta \rvert \lvert u \rvert} = 0$ of the sideband amplitude 
%%Therefore, the maximum sideband amplitude from the waveguide be expressed as:
%%
% \begin{equation}
%\bar{P}_{AS,max}=\frac{\kappa_{ex1}\kappa_{ex2}}{\kappa_1\kappa_2}
%\label{eq:s15}
%\end{equation}	
%
%

\section{Optomechanical Coupling Coefficient}

	\begin{figure}[b!]
	%	\begin{adjustwidth}{-1in}{-1in}
	\makebox[\textwidth][c]{\includegraphics[width=0.5\textwidth]{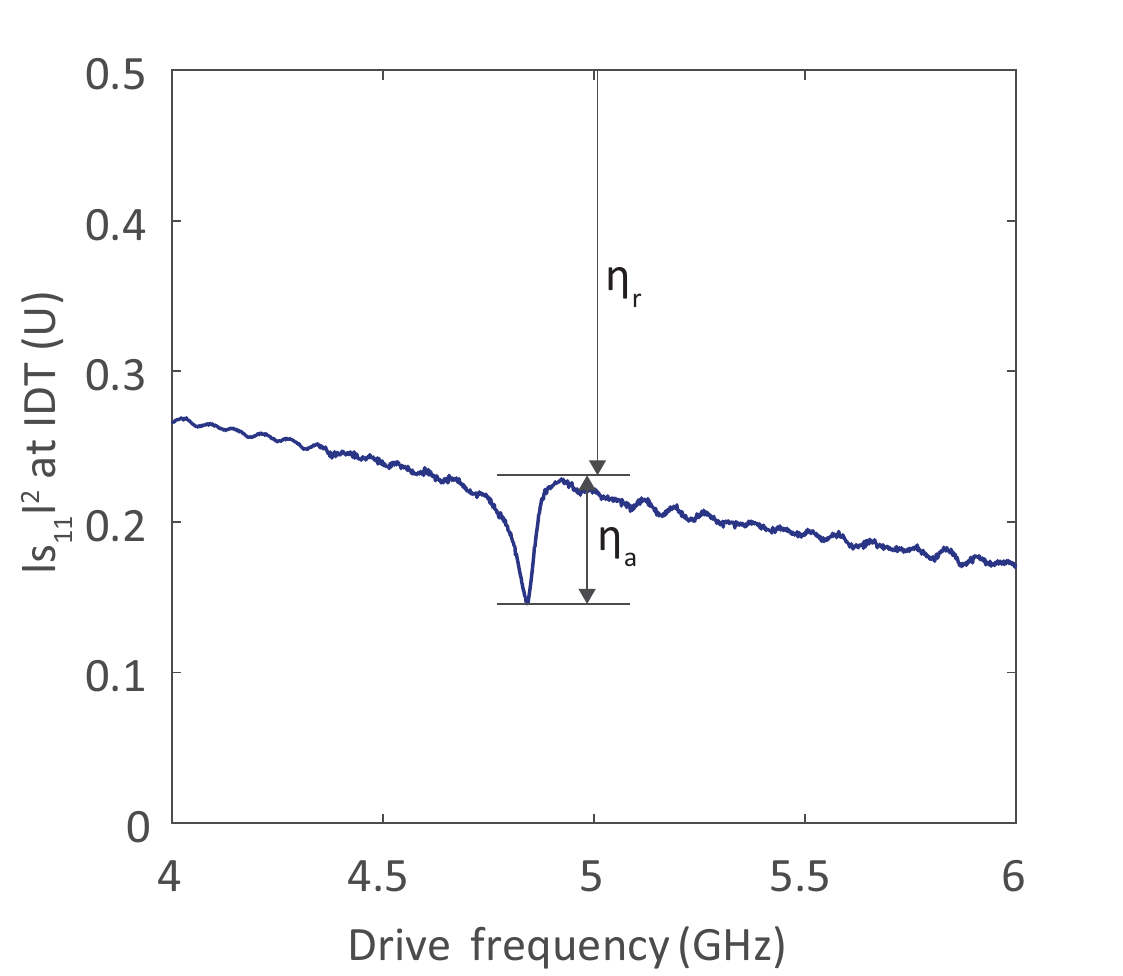}}
	\centering
	\caption{Reflection (s$_{11}$) measurement at IDT using a calibrated vector network analyzer (VNA). The S$_{0}$ Lamb acoustic wave is efficiently actuated at 4.82 GHz leading to a  reduction in the $s_{11}$ parameter (reflection). $\eta_a$ is the efficiency of conversion of electrical power to acoustic power. $\eta_r$ is the power loss at the IDT through parasitic capacitance. }
	\label{fig:WG1}
	%	\end{adjustwidth}
\end{figure}

	In order to quantify the intermodal optomechanical coupling coefficient ($\beta$) from the experimental data, we employ the model used in \cite{Li:15}. From the fitting of experimental data shown in Fig.5a (main manuscript), we obtain $G_{ph}=\lvert\beta\rvert \lvert u\rvert$ = 0.0589 GHz when 0 dBm of RF input power is supplied. Since, $\lvert\beta\rvert$ and $\lvert u\rvert$ cannot be independently determined from our fitting model, the displacement $u$ associated with the acoustic wave is estimated with the assistance of finite element simulation (COMSOL). The displacement associated with the acoustic pump for a given RF power is calculated using:
	
	\begin{align}
	u=\sqrt{\frac{2\pi \eta_aP_{RF}}{\gamma W \Omega}}
	\label{eq:15}
	\end{align}	
	
	\noindent where W is the IDT aperture, and $\gamma$ is the proportionality factor relating $u^2$ and acoustic energy $\gamma=\frac{1}{Wu^2} \int_{x,y,z=0}^{x,y,z=\Lambda,W,\infty} U(x,y,z)dxdydz$, $\Lambda$ is the wavelength of acoustic pump, and $U$ is the total mechanical energy density of mechanical mode. 
	
	The RF power transferred to the acoustic pump power is calculated using the $s_{11}$ parameter measurement with an electronic VNA. As shown in Fig.~\ref{eq:s4}, $\eta_a=~3.9\%$, implying that 3.9\% of the RF power is transferred to the acoustic power. $\gamma$ of the S$_0$ acoustic mode generated by the IDTs is calculated using FEM simulation (=$2.55\times 10^{11} J/m^3 $). From the fitting of Fig.~5a, the calculated $\beta$ is found to be 0.204 GHz/nm. 
	
	The calculated optomechanical coupling coefficient $\beta$ quantifies how efficiently the acoustic wave can couple the two optical modes. Note that the estimated displacement $u$ in the above  model is the acoustic response of the IDT, which is not the actual amplitude of the acoustic pump traveling in the waveguide. This is a limitation since there exist multiple sources of loss such as intrinsic acoustic loss of the material or reflections from the waveguide ridge, that do not permit a better estimate.

\section{Producing an Optical Isolator}

\begin{figure}[b!]
	%		\begin{adjustwidth}{-1in}{-1in}
	\makebox[\textwidth][c]{\includegraphics[width=0.9\textwidth]{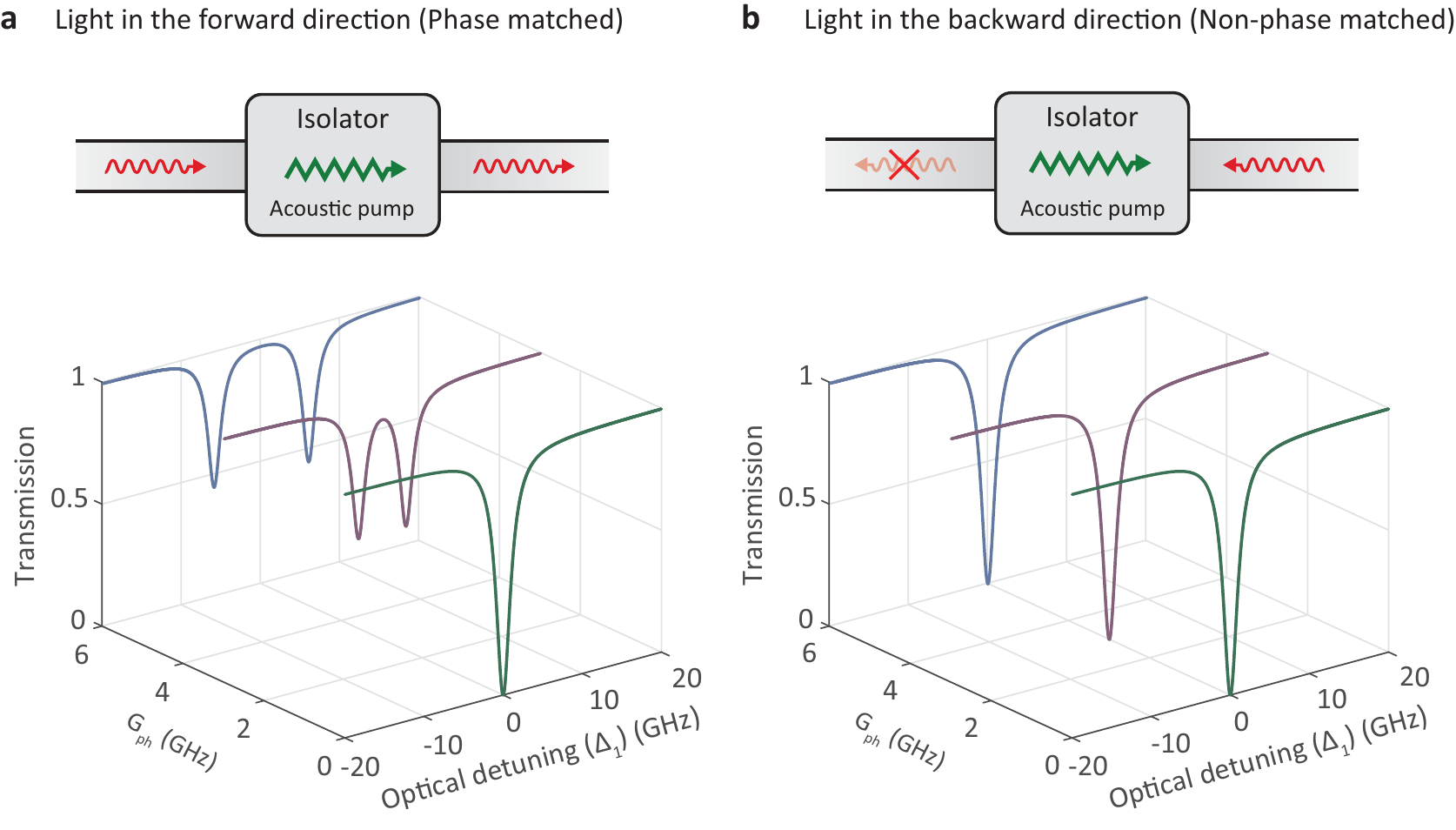}}
	\centering
	\caption{Conceptual schematic of optical isolator and 3D plot of light transmission spectrum respect to optomechanical coupling rate ($G_{ph}$) and TE$_{10}$ detuning ($\Delta_1$). (a) In the forward direction (phase matched direction), the optical mode is split and the transmission approaches 100 \% on resonance ($\Delta_1 = 0$) for large optomechanical coupling. (b) In the backward direction (non-phase matched direction) the optical mode shape remains the same so that light is absorbed by the resonator.}
	\label{fig:s6}
	%		\end{adjustwidth}
\end{figure}
	
	An optical isolator that blocks light in one direction but allows light transmission in the other direction is a canonical example of a non-reciprocal device (Fig.~\ref{fig:s6}). Here, we propose a theoretical model to build a magnetless optical isolator by simply changing few design parameters of our device. 
			
	Let us examine at the transmission spectrum of the TE$_{10}$ mode in the forward phase matched case (Eq.~\ref{eq:9}). We assume that the mechanical driving frequency and the frequency separation between the two optical modes are the same ($\omega_2 - \omega_1 = \Omega$) so that the intermodal scattering can most efficiently take place. We also assume that the sideband is in the fully resolved regime ($\Omega>>\kappa_1/2~\text{and}~\kappa_2/2 $ ). The power transmission from the waveguide in the forward direction can be now expressed as:
		\begin{align}
		\lvert s_{out} \rvert^2 = \lvert s_{in} - \sqrt{\kappa_{ex1}} a_{1,0} +\sqrt{\kappa_{ex2}} a_{2,+1} e^{-i\Omega t} \rvert^2
		\end{align}
	We note that $a_{2,0}$ and $a_{1,-1}$ terms approach zero when $\Delta_1 << \omega_2 - \omega_1$. Here, we make the TE$_{10}$ mode critically coupled to the waveguide ($\kappa_{ex1} = \kappa_1/2 =$ 1 GHz) so that there is no transmission through the waveguide. We also assume that the optical loss rate of TE$_{00}$ ($\kappa_2$) is 1 GHz. Using this equation, we plot the forward and backward transmission spectrum (Fig.~\ref{fig:s6}) with respect to the optical detuning of the laser from the TE$_{10}$ mode ($\Delta_1$).  Without optomechanical coupling ($G_{ph}$ = 0), the backward and forward transmissions exhibit identical Lorentzian shape transmission absorbing all the light on the resonance of the TE$_{10}$ mode ($\Delta_1=0$) as shown in Fig.~\ref{fig:s6}.  When the optomechanical coupling enters the strong coupling regime ($ G_{ph}\,>\,\sqrt{\kappa_1\kappa_2}/2$), the TE$_{10}$ resonance mode begins to split (Fig.~\ref{fig:s6}a). With sufficient acoustic power, the system becomes completely transparent at zero detuning only in the forward direction. In the backward direction, where the phase matching condition is not satisfied, the TE$_{10}$ resonance mode remains the same (Fig.~\ref{fig:s6}b) so that all the light is absorbed by the resonator.

\section{Producing Acousto-optic Frequency Shifter (AOFS)}
\begin{figure}[t!]
	%		\begin{adjustwidth}{-1in}{-1in}
	\makebox[\textwidth][c]{\includegraphics[width=0.45\textwidth]{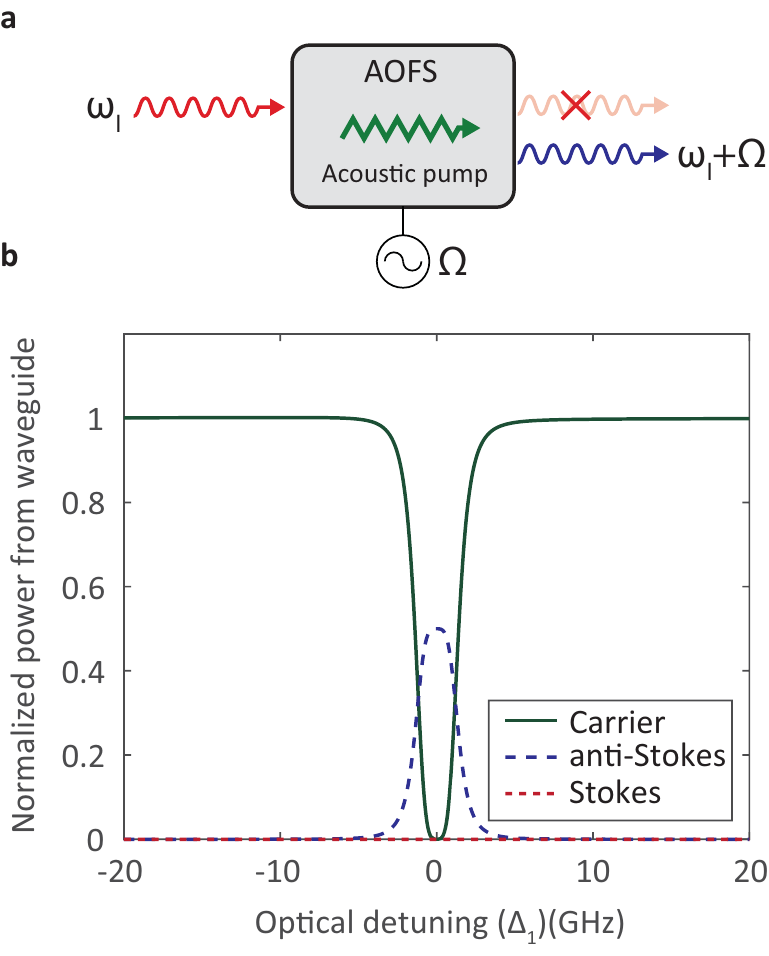}}
	\centering
	\caption{(a) Conceptual schematic of AOFS. (b) Example AOFS simulated based on the non-reciprocal modulator model. The carrier transmission is suppressed on resonance. 50\% of power is converted into the anti-Stokes sideband.}
	\label{fig:AOFS}
	%		\end{adjustwidth}
\end{figure}
%In this section, we would like to provide examples of applications that can be realized using the non-reciprocal modulator. 
%
%For simplicity, in the following, we assume two things; First, we assume that the frequency separation of the two optical resonance modes is very large compared to the line width of the two optical modes; Second, we assume that the mechanical frequency has the exact same value as the frequency separation between the two optical mode.
%
%
For an ideal AOFS, we desire carrier transmission to be attenuated so that the output of the device is only the frequency shifted light (Fig.~\ref{fig:AOFS}a). In this section, we theoretically provide an example of a chip-scale acousto-optic frequency shifter (AOFS) using our device.
%From Section S3, we can find the maximum sideband amplitude can be achieved when the following equation is satisfied.
%
% \begin{equation}
%\lvert G \rvert^2 = \frac{\kappa_1\kappa_2}{4} 
%\label{eq:16}
%\end{equation}		

Let us consider the transmission spectrum of TE$_{10}$ mode ($\Delta_1 << \omega_2 - \omega_1$) in the forward phase matched case. We assume that the two optical modes are fully resolved ($\omega_2-\omega_1 > \kappa_1/2 ~ \text{and} ~ \kappa_1/2$). The output from the waveguide can be expressed as:
\begin{align}
	\lvert s_{out} \rvert^2 = \lvert s_{in} - \sqrt{\kappa_{ex1}} \, a_{1,0} +\sqrt{\kappa_{ex2}} \,a_{2,+1} e^{-i\Omega t} \rvert^2
\end{align}
From Eq.~\ref{eq:7}a, we find that the transmission of the carrier laser vanishes ($s_{in}-\sqrt{\kappa_{ex1}}\,a_{1,0}~=~0$), when the following equation is satisfied.
\begin{equation}
\lvert G_{ph} \rvert^2 = \frac{\kappa_2}{2}\left(\kappa_{ex1}-\frac{\kappa_1}{2}\right)
\label{eq:17}
\end{equation} 
Therefore, the output from the waveguide is only an anti-Stokes sideband, which can be represented as:
\begin{equation}
	\lvert s_{out} \rvert^2 =\left| \frac{-iG_{ph}\sqrt{\kappa_{ex1}\kappa_{ex2}}}{(\kappa_2 /2 - i(\Delta_2+\Omega))(\kappa_1 /2 - i\Delta_1)+\lvert G_{ph} \rvert^2} s_{in}e^{-i\Omega t} \right|^2
	\label{eq:18}
\end{equation}
%By substituting Eq.~(\ref{eq:17}) into Eq.~(\ref{eq:16}),  
%\begin{equation}
%\kappa_{ex1}=\kappa_1
%\label{eq:18}
%\end{equation} 

\noindent From Eq.~\ref{eq:17}, we see that the carrier transmission can be fully suppressed for certain optomechanical coupling coefficient ($G_{ph}$) such that the resonator is over-coupled to the waveguide ($\kappa_{ex1}>\kappa_1/2$). In other words, when the combined photon loss rate induced by the scattering process and the intrinsic loss is the same as the external coupling rate, the optical mode is critically coupled to the waveguide. Based on the above equations, we calculate output spectrum of the TE$_{10}$ mode from the waveguide (Fig.~\ref{fig:AOFS}b). In this simulation, we assume that $\kappa_{ex1} = \kappa_{ex2} = 1.5~\text{GHz}$ and $\kappa_{1} =\kappa_2 = 2~\text{GHz}$. As shown in Figure~\ref{fig:s7}, when the acoustic power satisfying Eq.~\ref{eq:17} is applied to the system, the carrier is fully absorbed by the resonator and only the anti-Stokes light comes out from the waveguide.

\section{Racetrack Resonator Dimensions}

\begin{figure}[h!]
	%	\begin{adjustwidth}{-1in}{-1in}
	\makebox[\textwidth][c]{\includegraphics[width=0.6\textwidth]{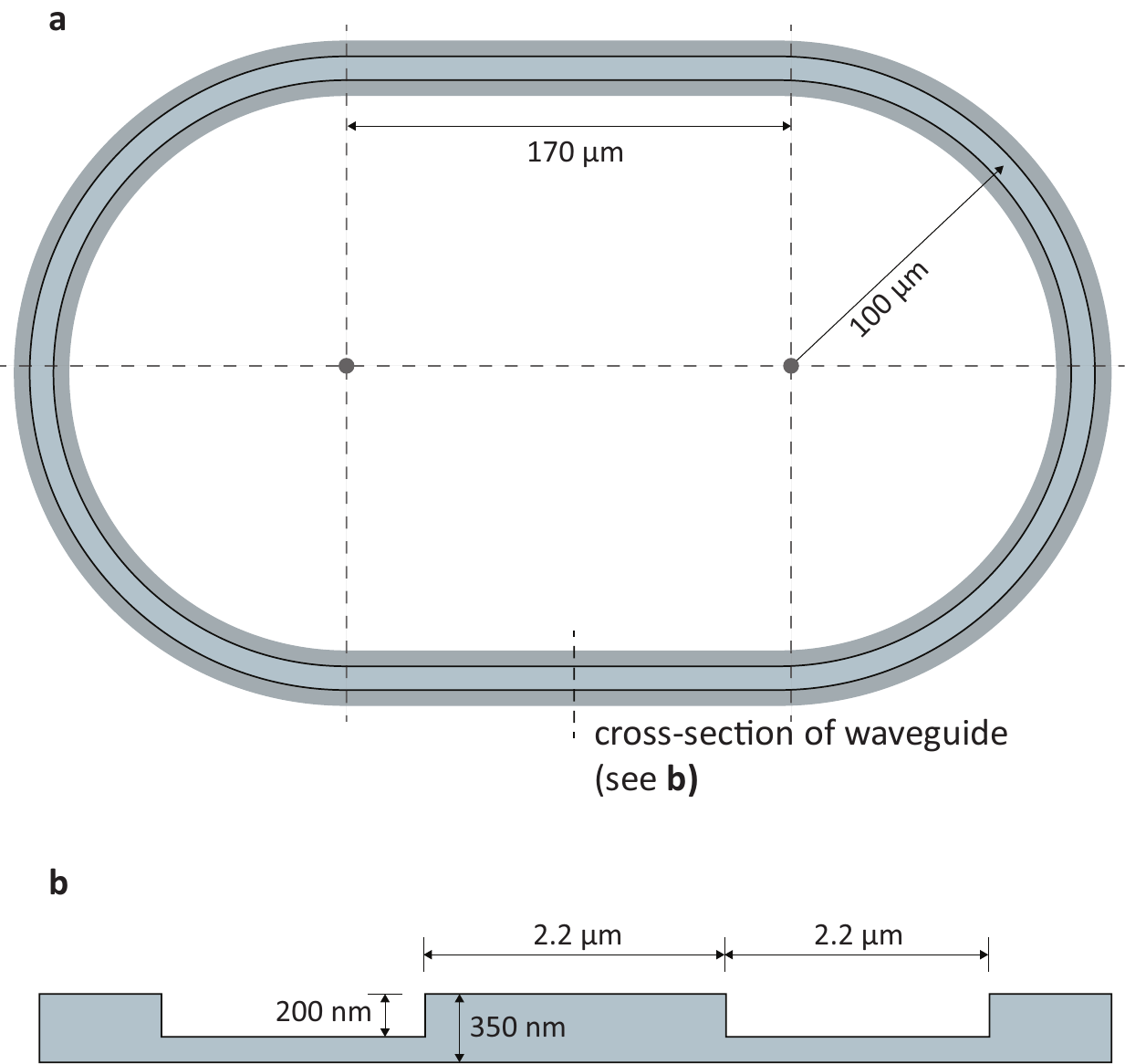}}
	\centering
	\caption{Dimensions of the racetrack optical resonator produced in this work that supports the required two optical modes.}
	\label{fig:s7}
	%	\end{adjustwidth}
\end{figure}

Specific geometry of our racetrack device is shown in Fig.~\ref{fig:s7}. The resonator is composed of a 2.2 $\mu$m width waveguide, having 170 $\mu$m linear regions and 100 $\mu$m radius curved regions. Thus, the total circumference of the racetrack resonator is 968 $\mu$m. The resonator is fabricated on 350 nm of AlN on silicon substrate, which is undercut to confine light and the S$_{0}$ Lamb acoustic wave in AlN only. The ridge for the racetrack is defined by 200 nm deep etches on either side. This leaves behind 150 nm of AlN such that the racetrack is mechanically supported by the rest of the substrate and is acoustically linked to it.

	%	This non-reciprocal modulator can be also used as optical isolator. Here, we propose two different approaches to build optical isolator. One method to build optical isolator is using non-reciprocal loss mechanism coming from the non-reciprocal scattering process. Let's consider a case where one optical mode is overcoupled. When the acoustic wave is applied to the system, additional loss of photon is induced due to the scattering processing. With the certain acoustic power, the external coupling rate and the loss rate becomes the same. in this case, the light is critically coupled to the resonator so that the transmission becomes 0. When the light enters in the opposite direction, the scattering does not occur since the phase matching condition is not satisfied. Therefore, optical transmission remains the same. With appropriate acoustic wave power, it is possible to achieve very high isolation contrast. The theoretical example of optical isolator using scattering process is shown in Fig. 5c. The parameter used in the calculation is the same with the parameter used in Section 5.1.

	%The optical isolator can be also built using non-reciprocal optical mode split. When high acoustic power is applied, the coupling between two optical modes become very strong. The optical modes start split after certain acoustic power. 

\newpage
\bibliographystyle{myIEEEtran}
\bibliography{sup_ref}